\documentclass[12pt,preprint]{aastex}
\def\beq{\begin{equation}}
\def\eeq{\end{equation}}
\def\ben{\begin{eqnarray}} 
\def\een{\end{eqnarray}}

\def\dunit{\,h^{-1}\,{\rm Mpc}}
\def\munit{\,h^{-1}\,M_{\odot}}

\def\mth{M_{\rm th}}
\def\uth{u_{\rm th}}

\def\jd{{\bf j}_{d}}
\def\js{{\bf j}_{s}}
\def\jg{{\bf j}_{g}}

\def\vor{{\bf w}}
\def\etm{{\bf e}_{t1}}
\def\eti{{\bf e}_{t2}}
\def\etn{{\bf e}_{t3}}
\def\evm{{\bf e}_{v1}}
\def\evi{{\bf e}_{v2}}
\def\evn{{\bf e}_{v3}}

\def\rv{r_{\rm vir}}
\def\rs{R_{\rm s}}
\def\rf{R_{f}}
\def\rh{r_{1/2\star}}
\def\rw{R_{\rm w}}
\def\im{\iota}

\usepackage{xcolor}

\begin{document}
\title{Radius-Dependent Spin Transition of Dark Matter Halos}
\author{Jun-Sung Moon$^{1,2}$ and Jounghun Lee$^{1}$}
\affil{$^1$Astronomy Program, Department of Physics and Astronomy,
Seoul National University, Seoul 08826, Republic of Korea
\email{jsmoon.astro@gmail.com, jounghun@astro.snu.ac.kr}}
\affil{$^2$Research Institute of Basic Sciences, Seoul National University, Seoul 08826, Republic of Korea}


\begin{abstract}
A numerical detection of the radius-dependent spin transition of dark matter halos is reported. Analyzing the data from the IllustrisTNG simulations, 
we measure the halo spin vectors at several inner radii within the virial boundaries and investigate their orientations 
in the principal frames of the tidal and velocity shear fields, called the Tweb and Vweb, respectively. 
The halo spin vectors in the high-mass section exhibit a transition from the Tweb intermediate to major principal axes as they 
are measured at more inner radii, which holds for both of the dark matter and baryonic components. 
The radius threshold at which the transition occurs depends on the smoothing scale, $\rf$, becoming larger as $\rf$ decreases. 
For the case of the Vweb, the occurrence of the radius-dependent spin transition is witnessed only when $\rf\ge 1\dunit$. 
Repeating the same analysis but with the vorticity vectors, we reveal a critical difference from the spins. The vorticity vectors are always 
perpendicular to the Tweb (Vweb) major principal axes, regardless of $\rf$, which indicates that the halo inner spins are not strongly affected 
by the generation of vorticity.
It is also shown that the halo spins, as well as the Tweb (Vweb) principal axes, have more directional coherence over a wide range of radial distances 
in the regions where the vorticity vectors have higher magnitudes. The physical interpretations and implications of our results are discussed. 
\end{abstract}
\section{Introduction}\label{sec:intro}

Numerous N-body simulations have so far confirmed that the halo spin vectors tend to be intrinsically aligned with the cosmic web 
\citep[see][for a review]{align_review1,align_review2}. 
It is believed that the initial tidal interactions between the protohalos and the surrounding matter distribution originate the intrinsic spin 
alignments \citep{whi84,lp00,LE07} 
and that the nonlinear processes like hierarchical merging in due subsequence should modify the alignment 
tendency and strength \citep{ara-etal07,hah-etal07,pic-etal11,cod-etal12}. 
The unique aspect of the intrinsic spin alignments of dark matter (DM) halos is that the preferred spin orientations show a transition  
from being parallel to being perpendicular to the hosting filaments as the halo mass increases 
\citep[e.g.,][]{ara-etal07,hah-etal07,paz-etal08,cod-etal12,TL13,tem-etal13,tro-etal13,lib-etal13a,AY14,dub-etal14,for-etal14,cod-etal15a,
cod-etal15b,hir-etal17,cod-etal18,gan-etal18,wan-etal18,gan-etal19,kra-etal20}. 

This mass-dependent spin transition of DM halos was conventionally interpreted as a manifestation of the merging effect on the spin orientations 
\citep{ara-etal07,hah-etal07,pic-etal11}.  
For the case of the high-mass halos which form through frequent merging events that preferentially occur along the filaments, their spin axes are 
driven to be aligned with the directions perpendicular to the filaments. Meanwhile, for the case of the low-mass halos that undergo less frequent mergers, 
they retain the initial tidal memory being aligned with the directions along the filaments \citep[e.g.,][]{cod-etal12,tro-etal13,dub-etal14,cod-etal18,gan-etal18,kro-etal19}. 

Another nonlinear process that can severely affect the evolution of the halo spin orientations is the generation of vorticity. 
In the linear regime, the peculiar velocity field is curl-free and proportional to the gradient of the perturbation potential. 
As it evolves, however, the nonlinearly evolved velocity field develops a curl mode, i.e., the vorticity \citep{PB99,PS09,kit-etal12} on a small scale,  
which can affect the halo angular momentum, amplifying its magnitude and reorienting its direction. 
Showing by N-body simulations that the vorticity vectors measured at the halo positions exhibit a strong alignment with the halo spins, \citet{lib-etal13b} 
claimed that the generation of vorticity in the nonlinear regime is largely responsible for the occurrence of the spin transition phenomenon as well as 
for the intrinsic spin alignments of DM halos with the cosmic web in the nonlinear regime \citep[see also][and references therein]{lib-etal14}. 

In this picture, it is naturally expected that the variation of the tendencies and strengths of the intrinsic spin alignments with radial distance, say, 
{\it the spin alignment profile}, should be a good indicator of the evolutionary processes that the halos experience.  
At inner radii, it may reflect the nonlinear effects including violent relaxation and generation of vorticity, while at outer radii, it should contain the memory of the 
latest mergers and infalls along the cosmic web.
Complementing the mass density profile of DM halos that is most commonly used to describe the halo internal structure \citep{nfw96},  
the spin alignment profiles of DM halos may be useful to understand their merging history and internal dynamics and to relate them with the cosmic web.  

Besides, what can be observed from real galaxies is the stellar spin axes that are usually measured at inner radii much smaller than the virial boundaries 
\citep{rom-etal03,ems-etal07,coc-etal09,cor-etal16,wel-etal20}. For a proper test of any theoretical predictions against observations, it is urged to find the 
halo spin alignment profile which can link theoretically predictable spin vectors at virial boundaries with observationally measurable spin directions at inner 
radii. In our prior work of \citet{lee-etal21}, it was found from high-resolution hydrodynamical simulations that the stellar spin vectors of 
massive galactic halos exhibit a {\it peculiar} tendency of being aligned with the directions of maximum matter compression, to which the DM spin vectors are 
always perpendicular, even though the stellar and DM spin vectors exhibit fairly strong alignments with each other. 
It was originally suspected that non-gravitational baryonic feedbacks after merging events might be responsible for this peculiar tidal 
connection of the galaxy stellar spin axes \citep{lee-etal21,lee-etal22}. In the follow-up works, however, it was revealed that the peculiar alignment 
tendency of the stellar spin vectors is likely to be established during the quiescent period when no merging events disturb the galactic halos \citep{LM22}. 

A key difference, however, existed between the DM and stellar components in the way that their spin directions are measured with respect to the cosmic web. 
While the former was measured at the halo virial boundary as usual,  twice the stellar half-mass radius, $2\rh$, was used for the measurements of the 
latter, given the observational limitations \citep{rom-etal03,ems-etal07,coc-etal09,cor-etal16,wel-etal20}. 
The consequential issues regarding the halo spin alignment profile are the followings. First, will the DM spin axes of the halos, if measured at $2\rh$, 
also exhibit similar peculiar alignments with the directions of maximum matter compression? Second, is the type of the halo spin transition dependent on the 
radii at which the spin vectors are measured? If so, what will be the transition radius threshold? Third, does the generation of vorticity have anything to do with 
the peculiar alignments of the stellar spin vectors?

In this paper, we will address all of these issues by analyzing the high-resolution N-body and hydrodynamical simulations. 
For a more comprehensive analysis, we will use two different web-identification algorithms, since the tendency and strength of the intrinsic spin alignments of the DM 
halos are known to depend on how to define the cosmic web \citep[e.g.,][]{for-etal14}. The upcoming sections contain the following contents. 
In Section \ref{sec:data}, the simulation data are briefly described, and the numerical analysis is fully laid out. In Section \ref{sec:result}, the results of the analysis and 
our physical interpretations of them are presented. In Section \ref{sec:sum}, a concise summary of the main results is provided, and a final conclusion is drawn. 
Throughout this paper, we will assume a flat universe whose energy densities are dominantly contributed by the cosmological constant $\Lambda$ and 
cold DM at the present epoch.

\section{Data and Numerical Analysis}\label{sec:data}

The data on which our analysis is based are from two cosmological simulations: a DM-only N-body simulation called the TNG300-1-Dark and a hydrodynamic 
simulation called the TNG300-1, both of which were performed on a periodic box of volume $250\,h^{-3}$Mpc$^{3}$ for a Planck $\Lambda$CDM 
cosmology \citep{planck16}, appertaining to a suite of the IllustrisTNG project \citep{tngintro1, tngintro2, tngintro3, tngintro4, tngintro5, illustris19}. 
The TNG300-1-Dark simulation contains only $2500^{3}$ DM particles of mass resolution $m_{\rm d} = 7.0\times 10^{7}\,M_{\odot}$, while the 
TNG300-1 simulation contains not only DM particles of resolution $m_{\rm d} = 5.9\times 10^{7}\,M_{\odot}$ but also an equal number of baryonic gas cells of initial 
resolution $m_{\rm b}=1.1\times 10^{7}\,M_{\odot}$, whose evolution was tracked by the Arepo code \citep{arepo} that is capable of incorporating the essential hydrodynamical 
processes.We refer the readers to the IllustrisTNG web page\footnote{https://www.tng-project.org/data/} for full information on the simulations. 

The halo catalogs from each of the two simulations contain the bound groups and their substructures identified by the friends-of-friends 
(FoF) and Subfind algorithms \citep{subfind}, respectively.  From here on, we will refer to the substructures of the FoF groups as the halos. 
We first investigate how many particles each halo from the TNG300-1-Dark simulation at $z=0$ has within a certain radius, $\rs$, from the center.  
The value of $\rs$ is chosen to be smaller than the virial boundary of each halo, $\rv$, within which the spherically averaged mass density exceeds $200$ times 
the critical density. Then, we select only those halos consisting of more than $300$ DM particles within $\rs$ to calculate the angular momentum vector 
at an inner radius, ${\bf J}(\rs)$: 
\begin{equation}
\label{eqn:dm_spin}
{\bf J} (\rs) = \sum_{\alpha=1}^{n_{d}}{m}_{d}\,[({\bf x}_{d,\alpha}-{\bf x}_{c})\times({\bf v}_{d,\alpha}-{\bf v}_{c})]\, ,\\
\end{equation}
where $n_{d}$ is the number of DM particles within $\rs$, ${\bf x}_{d,\alpha} = (x_{d,\alpha\,i})$ and ${\bf v}_{d,\alpha}= (v_{d,\alpha\,i})$ are the comoving position 
and peculiar velocity of the $\alpha$th DM particle within $\rs$,  respectively, while ${\bf x}_{c}=(x_{c,i})$ and ${\bf v}_{c}= (v_{c,i})$ are the positions and velocities 
of the halo center, respectively.  We exclude those halos with $n_{d}\leq300$ on the ground that the presence of shot noise would significantly contaminate the calculation 
of ${\bf J}(\rs)$ if $n_{d}\leq300$ \citep{bet-etal07}.  Hereafter, the unit angular momentum vector, $\jd (\rs) \equiv {\bf J}(\rs)/\vert{\bf J}(\rs)\vert$, 
will be called the halo inner (virial) spin vector if $\rs<\rv$ ($\rs=\rv$).    

For each halo from the TNG300-1 simulations, we calculate the inner spin vectors of the DM, non-stellar gas, and stellar components separately, 
denoted by $\jd (\rs)$, $\jg (\rs)$, and $\js (\rs)$, respectively, in a very similar manner. Given that for real galaxies it is usually possible to measure 
their stellar spin vectors only within $2\rh$ due to the observational limitation \citep{rom-etal03,ems-etal07,coc-etal09,cor-etal16,wel-etal20},  
we compute $\jd (\rs)$, $\jg (\rs)$ and $\js (\rs)$ at two different inner radii, $\rs=\rh$ and $\rs=2\rh$. 
It is worth mentioning here that we apply the particle number cut, $300$, separately to each component. For example, for the computation of $\js (\rs)$, 
we select only those halos consisting of more than $300$ {\it stellar} particles. 

From the full snapshot data of the two simulations at $z=0$ released by the IllustrisTNG project, we construct three different fields: the tidal, velocity shear, 
and vorticity fields smoothed by a Gaussian kernel with a filtering radius of $\rf$ on the $512^{3}$ grid points. For the case of the TNG300-1 simulation, 
all of the DM, non-stellar gas, and stellar particles are used to construct these three fields, while for the other case only the DM particles are used. 
For the construction of the tidal fields, we follow the same procedure\footnote{In the previous works of \citet{lee-etal21} and \citet{lee-etal22}, the tidal 
fields are reconstructed from the spatial distributions of DM halos. In the current work, it is from the particle snapshot.} 
as described in \citet{lee-etal21} and \citet{lee-etal22}. 
We first construct the real-space density contrast field, $\delta ({\bf x})$, out of the spatial distributions of all particles with the help of the cloud-in-cell method and 
then carry out a Fourier transformation of it, $\delta({\bf k})$, to compute the Gaussian filtered Fourier-space tidal field, 
${\bf T}({\bf k})=\left[T_{ij}({\bf k})\right]$ with $T_{ij}({\bf k})=[k_{i}k_{j}\delta({\bf k})/\vert{\bf k}\vert^2]\exp(-\vert{\bf k}\vert^2\rf^2/2)$.  
Finally, we construct the real-space tidal field, ${\bf T}({\bf x})=\left[T_{ij}({\bf x})\right]$, through an inverse Fourier transformation and find their three eigenvalues 
and corresponding orthonormal eigenvectors, ${\bf e}_{t}=(e_{ti})$, through the similarity transformation at the grids where the selected halos are located. 
The directions of maximum (minimum) matter compression at the halo positions are often regarded as being parallel to $\etm$ ($\etn$) 
corresponding to the  largest (smallest) eigenvalue of ${\bf T}$. Meanwhile, the linear tidal torque theory predicts initial alignments of the halo spin orientations 
with $\eti$. Hereafter, the orientations of the cosmic web defined by the axes of $\etm$, $\eti$, and $\etn$ will be referred to as {\it the Tweb major, intermediate, and 
minor principal axes}, as in \citet{for-etal14}, respectively. 

\begin{figure}[ht]
\centering
\includegraphics[width=\textwidth]{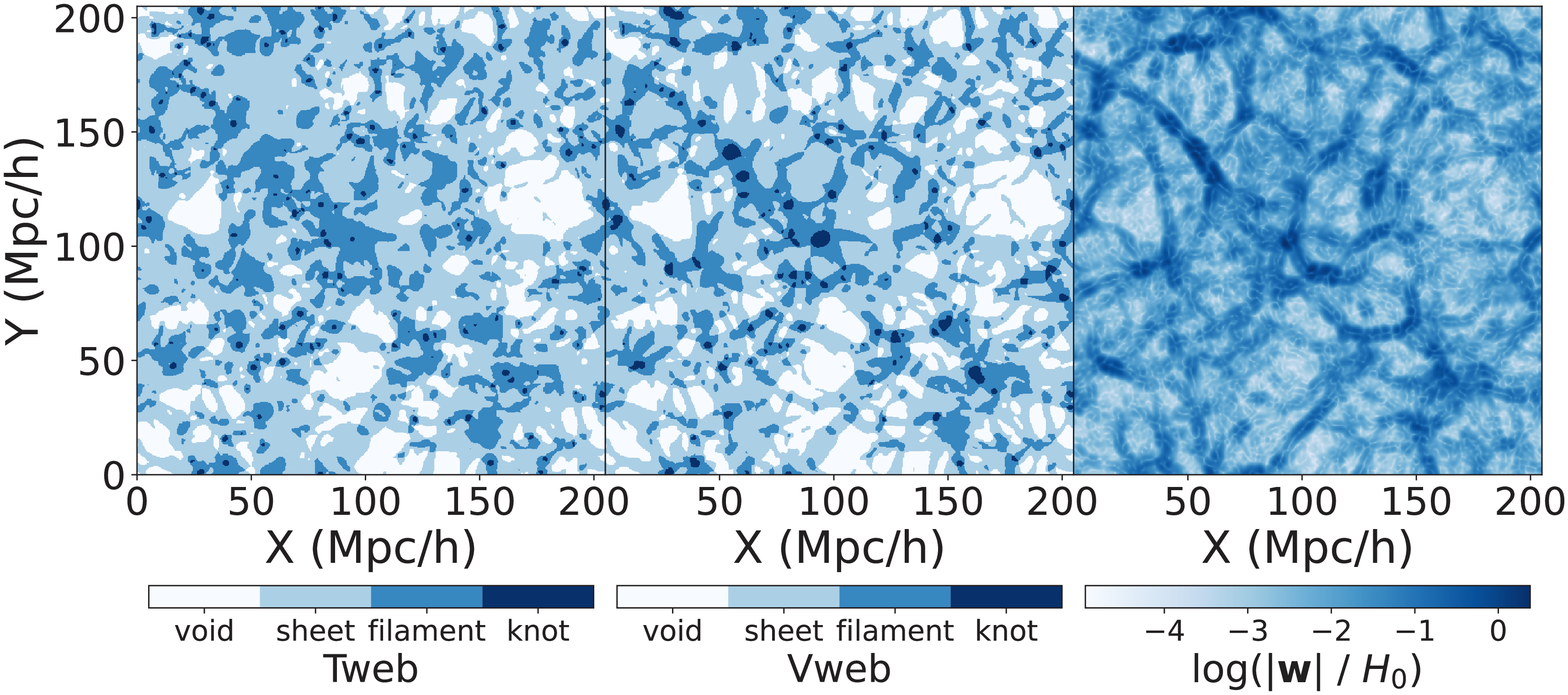}
\caption{Four types of the cosmic web identified by the eigenvalue signs of the Tweb (left panel) and Vweb (middle panel) 
smoothed on the scale of $\rf=1\dunit$, and the magnitudes of the vorticity field (right panel) in the two dimensional 
space projected onto the Cartesian $z$-axis from the TNG300-1-Dark simulation.}
\label{fig:snap}
\end{figure}

Taking the exactly same prescription given by \citet{lib-etal13a} and \citet{lib-etal14},  we also construct the velocity shear ${\bf\Sigma}=(\Sigma_{ij})$ and 
vorticity fields $\vor = (w_{i})$ by separating the deformation tensor field into a symmetric and an anti-symmetric term. Applying the cloud-in-cell 
method to the comoving peculiar velocities of all particles, we first calculate the mean peculiar velocity field, ${\bf v}({\bf x})=(v_{i})$, on the $512^{3}$ grids. 
Then, we find a Fourier space velocity field, ${\bf v}({\bf k})$, to obtain a Gaussian filtered Fourier-space deformation tensor field, $\xi_{ij}({\bf k})$, as 
$\xi_{ij}({\bf k})\equiv \im k_{j}v_{i}({\bf k})\exp(-\vert{\bf k}\vert^2\rf^2/2)$, where $\im$ is the imaginary unit. As prescribed in \citet{lib-etal14}, we express $\xi_{ij}({\bf k})$ 
as a sum of its symmetric and anti-symmetric terms, ${\bf\Sigma}=(\Sigma_{ij})$ and ${\bf\Gamma}=(\Gamma_{ij})$, respectively, 
\begin{eqnarray}
\label{eqn:xi_def}
\xi_{ij}({\bf k}) &=& {H(z)}\left[-\Sigma_{ij}({\bf k}) + \Gamma_{ij}({\bf k})\right] \\
\label{eqn:sym_antisym}
&\equiv& \frac{1}{2}\left[\im k_{j}v_{i}({\bf k}) + \im k_{i}v_{j}({\bf k})\right]e^{-\vert{\bf k}\vert^2\rf^2/2} + \frac{1}{2}\left[\im k_{j}v_{i}({\bf k}) - \im k_{i}v_{j}({\bf k})\right]e^{-\vert{\bf k}\vert^2\rf^2/2}\,,
\end{eqnarray}
where $H(z)$ is the Hubble constant at a given redshift $z$. 
The Fourier-space vorticity field, $\vor({\bf k})$ is then determined from the anti-symmetric tensor, $\Gamma_{ij}({\bf k})$, as 
$w_{i} = -\epsilon_{ijk}\Gamma_{jk}H(z)$ where $\epsilon_{ijk}$ is the anti-symmetric Levi-Civita symbol.
The inverse Fourier transformations of ${\bf \Sigma}({\bf k})$ and ${\bf w}({\bf k})$ finally lead us 
to obtain the real-space Gaussian filtered velocity shear and vorticity fields, ${\bf \Sigma}({\bf x})$ and ${\bf w}({\bf x})$, respectively. 
The orthonormal eigenvectors of ${\bf \Sigma}({\bf x})$ at the halo positions are also found via the similarity transformation and denoted by $\evm$, $\evi$, and $\evn$ 
corresponding to the largest, second largest, and smallest eigenvalues, respectively. The cosmic web whose orientations are parallel to $\{\evm$, $\evi$, $\evn\}$ is 
called the Vweb \citep[e.g.,][]{lib-etal13a,for-etal14}.  
The left (middle) panel of Figure \ref{fig:snap} plots the four types of the cosmic web, namely, knots, filaments, sheets, 
and voids identified by the signs of the Tweb (Vweb) eigenvalues \citep{hah-etal07} on the scale of $1\dunit$, while its right panel depicts the magnitudes of the vorticity vectors 
on the same scale. These plots are all made in the two dimensional space projected onto the Cartesian $z$-axis from the TNG300-1-Dark simulation. 

To investigate the radius dependence of the halo spin alignments with the cosmic web from the TNG300-1-Dark simulation, we basically calculate the 
following quantities as a function of halo total mass, $M_{\rm h}$ (i.e., the sum of the masses of all member particles) whose range is split into small bins, 
\begin{equation}
\{\langle\vert\jd(\rs)\cdot{\bf e}_{ti}(\rf)\vert\rangle, \langle\vert\jd(\rs)\cdot{\bf e}_{vi}(\rf)\vert\rangle | i\in \{1,2,3\},\rs\in \{\rv,\rv/2,\rv/4,\rv/8\}\}\, ,
\end{equation}
for three different cases of the smoothing scale of $\rf/(h^{-1}{\rm Mpc})\in \{0.5,\ 1,\ 2\}$. Here, the ensemble average is taken over the halos whose masses fall in 
the same mass bin.  The associated errors are also calculated as one standard deviations in the mean values at each mass bin. 
The halo inner spins, $\jd(\rs)$, will be described as being preferentially aligned with the $i$th principal axis of the Tweb on the scale of $\rf$ if the 
following conditions are satisfied: (i) The ensemble average, $\langle\vert\jd(\rs)\cdot{\bf e}_{i}(\rf)\vert\rangle$, is higher than $0.5$ with statistical 
significance.  (ii) It is also higher than $\langle\vert\jd(\rs)\cdot{\bf e}_{j}(\rf)\vert\rangle$ and $\langle\vert\jd(\rs)\cdot{\bf e}_{k}(\rf)\vert\rangle$ with 
statistical significance where $j\ne i$ and $k\ne i$. These criteria will be consistently used to test the existence of the alignments of the other vectors 
throughout this paper. 

To explore how the alignment tendency of the halo inner spins differs among the DM, gas and stellar components from the TNG300-1 simulations, 
we also separately calculate 
\begin{eqnarray}
&&\{\langle\vert\jd(\rs)\cdot{\bf e}_{ti}(\rf)\vert\rangle, \langle\vert\jd(\rs)\cdot{\bf e}_{vi}(\rf)\vert\rangle | i\in \{1,2,3\},\rs\in \{\rh,2\rh\}\}\, , \\
&&\{\langle\vert\jg(\rs)\cdot{\bf e}_{ti}(\rf)\vert\rangle, \langle\vert\jg(\rs)\cdot{\bf e}_{vi}(\rf)\vert\rangle | i\in \{1,2,3\},\rs\in \{\rh,2\rh\}\}\, , \\
&&\{\langle\vert\js(\rs)\cdot{\bf e}_{ti}(\rf)\vert\rangle, \langle\vert\js(\rs)\cdot{\bf e}_{vi}(\rf)\vert\rangle | i\in \{1,2,3\},\rs\in \{\rh,2\rh\}\}\, ,
\end{eqnarray}
for the same three cases of $\rf$. To see how the vorticity vectors are aligned with the halo inner spins from the TNG300-1-Dark simulation, we calculate 
\begin{equation}
\{\langle\vert\jd(\rs)\cdot{\vor}(\rw)\vert\rangle | \rs\in \{\rv,\rv/2,\rv/4,\rv/8\}, \rw\in\{4\rv\}\}\, .
\end{equation}
Here ${\vor}(\rw)$ represents the vorticity vectors at each mass bin smoothed on a scale of $\rw=4\rv$ with corresponding 
mean virial radii $\rv$. Unlike the tidal and velocity shear fields that are smoothed on a fixed scale of $\rf$, the vorticity field is smoothed on a varying scale,  
given that the vorticity effect on the halos with mass $M_{h}$ is known to rapidly fade away on the scales beyond four times the corresponding virial radii 
$\rv$ \citep{lib-etal13a,lib-etal13b,lib-etal14}.  

To investigate how the presence of baryons affects the spin-vorticity alignments, we calculate 
\begin{eqnarray}
&&\{\langle\vert\jd(\rs)\cdot{\vor}(\rw)\vert\rangle | \rs\in \{\rh,2\rh\}, \rw\in \{4\rv\},\}\, , \\
&&\{\langle\vert\jg(\rs)\cdot{\vor}(\rw)\vert\rangle | \rs\in \{\rh,2\rh\}, \rw\in \{4\rv\},\}\, , \\
&&\{\langle\vert\js(\rs)\cdot{\vor}(\rw)\vert\rangle | \rs\in \{\rh,2\rh\}, \rw\in \{4\rv\},\}\, , 
\end{eqnarray}
from the TNG300-1 simulations. To see if the vorticity vectors also exhibit a similar alignment in the principal frame of the Tweb (Vweb) even at inner radii 
to the spin vectors, we also calculate
\begin{equation}
\{\langle\vert\vor(4\rv)\cdot{\bf e}_{ti}(\rf)\vert\rangle, \langle\vert\vor(4\rv)\cdot{\bf e}_{vi}(\rf)\vert\rangle | i\in \{1,2,3\}\, , 
\rf/(h^{-1}{\rm Mpc})\in \{0.5,\ 1,\ 2\}\}\, , 
\end{equation}
from both of the TNG300-1-Dark and TNG300-1 simulations. 
 
To explore the effect of vorticity on the strengths of the alignments between the halo inner and virial spins from the TNG300-1-Dark simulation, 
we also divide the selected halos into four subsamples according to their values of $\jd(\rv)\cdot\jd(\rv/8)$ and control them to share the identical 
mass and density joint distributions. We take the ensemble average of $\log\vert \vor\vert/H_{0}$ over each of the controlled samples. 
If $\langle\log\vert \vor\vert/H_{0}\rangle$ turns out to significantly differ among the four subsamples, then it can be attributed not to the differences in $M_{h}$ 
and $\delta$ but only to the difference in $\jd(\rv)\cdot\jd(\rv/8)$. In other words, it would confirm the existence of the net vorticity effect on the alignment strengths 
of the halo spins at different radii from the TNG300-1-Dark. 
The alignments between the small-scale Tweb and the large-scale Vweb major principal axes,  $\vert\etm(\rf=0.5\dunit)\cdot\evm(\rf=2\dunit)\vert$, are also 
used to create another set of four controlled subsamples and then to investigate if and how $\langle\log\vert \vor\vert/H_{0}\rangle$  varies among them. 
We also repeat the same calculations but from the TNG300-1 simulation to see if the presence of baryons affects the relation between 
the vorticity magnitude and the spin alignments, if any. 

\section{Results and Physical Interpretations}\label{sec:result}
\subsection{Inner Spin Alignments with the Cosmic Web}\label{sec:tweb}

\begin{figure}[ht]
\centering
\includegraphics[height=18cm,width=12cm]{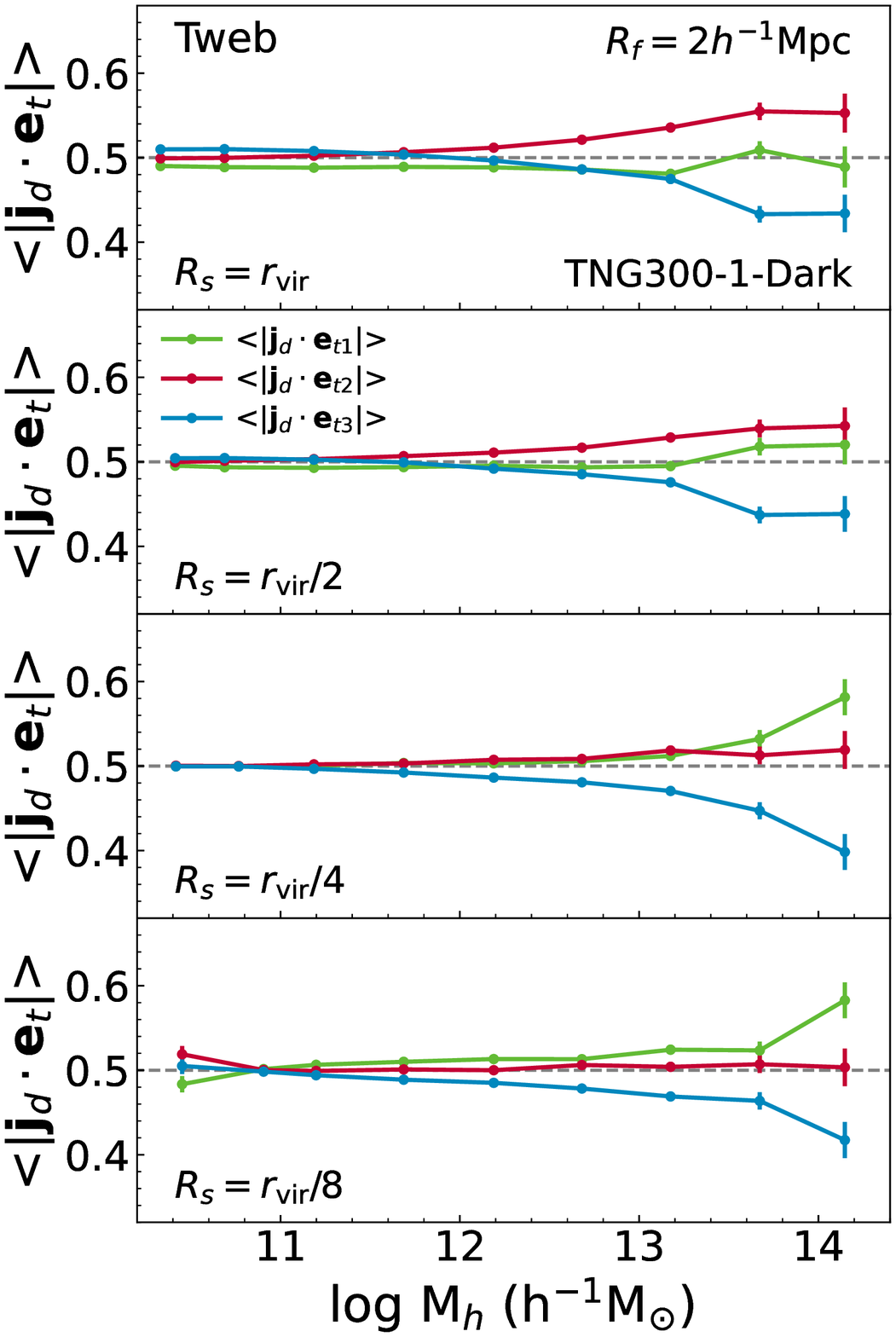}
\caption{Mean absolute values of the cosines of the angles between the halo spin vectors measured at 
$\rv$, $\rv/2$, $\rv/4$ and $\rv/8$ (from top to bottom panels, respectively) and three principal axes of the 
tidal fields smoothed on the scales of $\rf=2\dunit$ as a function of halo mass at $z=0$ from the TNG300-1-Dark simulation.}
\label{fig:je_m_rf2}
\end{figure}

Figure \ref{fig:je_m_rf2} shows how the mass variations of the alignment tendencies between the halo inner spins and the Tweb principal axes depend 
on the radial distances, $\rs$,  for the case of $R_{f}=2\dunit$, from the TNG300-1-Dark simulation at $z=0$. 
As can be seen, at $\rs=\rv$, we reproduce the well known mass-dependent spin transition phenomenon: 
the preferred directions of the halo virial spins transit from the Tweb intermediate to minor principal axes as the halo mass decreases below a threshold  
\footnote{\citet{lee-etal20} proposed a very refined rigorous way to determine the threshold mass at which the spin transition 
occurs based on the Kolmogorov–Smirnov test. In the current analysis, however, we do not adopt this rigorous method because it is not the main focus of our work 
to accurately determine the transition threshold. Rather, we use the simple conventional criterion mentioned in Section \ref{sec:data}.} 
around $M_{h}\approx 10^{12}\munit$ \citep[e.g.,][]{ara-etal07,hah-etal07,cod-etal12,TL13,AY14,for-etal14,cod-etal18,lee-etal21}. 
This type of spin transition was dubbed the type II spin transition by \citet{lee-etal21} in their attempt to differentiate it from the type I spin transition between 
$\etm$ and $\etn$ exhibited by the stellar spin vectors.  
Note also that the halo virial spins seem to be perpendicular to the Tweb major principal axes over almost the entire mass range, which is consistent with the results of 
the previous works \citep{lib-etal14,for-etal14,lee-etal21}. 

At inner radii $\rs<\rv$, however, we witness quite different phenomena.  At $\rs=\rv/2$, the inner spins of the massive halos with $M_{h}\gtrsim 10^{13}\munit$ 
are still strongly aligned with the Tweb intermediate principal axes. Although the strength of the $\jd$--$\eti$ alignment decreases as the halo mass decreases similar 
to the case of $\rs=\rv$, no significant signal of the mass-dependent type II spin transition is found for this case.
At $\rs=\rv/4$, the inner spins of the halos with $M_{h}\gtrsim 10^{13}\munit$ exhibit the {\it peculiar} alignments with the Tweb major principal axes, 
being random with respect to the intermediate and perpendicular to the minor principal axes.  In spite of the rapid decrease of the $\jd$--$\etm$ alignment strength 
with the decrement of $M_{h}$, no significant signal of the occurrence of the mass-dependent halo spin transition is found.  
At $\rs=\rv/8$, the alignments of the halo inner spins with the Tweb major principal axes are witnessed in the larger mass range of $M_{h}\gtrsim 10^{12}\munit$. 
These results clearly demonstrate that the halo spin transition occurs not only in a mass-dependent way but also in a radius-dependent way 
and that the radius threshold for the spin transition between $\eti$ and $\etm$ is not universal but dependent upon the halo virial radii.  
The lower mass limit, $\mth$, above which the $\jd(\rs)$--$\etm$ alignment tendency is found becomes smaller as the ratio $\rs/\rv$ decreases. 
In other words, the alignments between the halo inner spins and the Tweb major principal axes can be found even in the lower mass section if the inner spins are 
measured at more inner radii. 

\begin{figure}[ht]
\centering
\includegraphics[height=18cm,width=12cm]{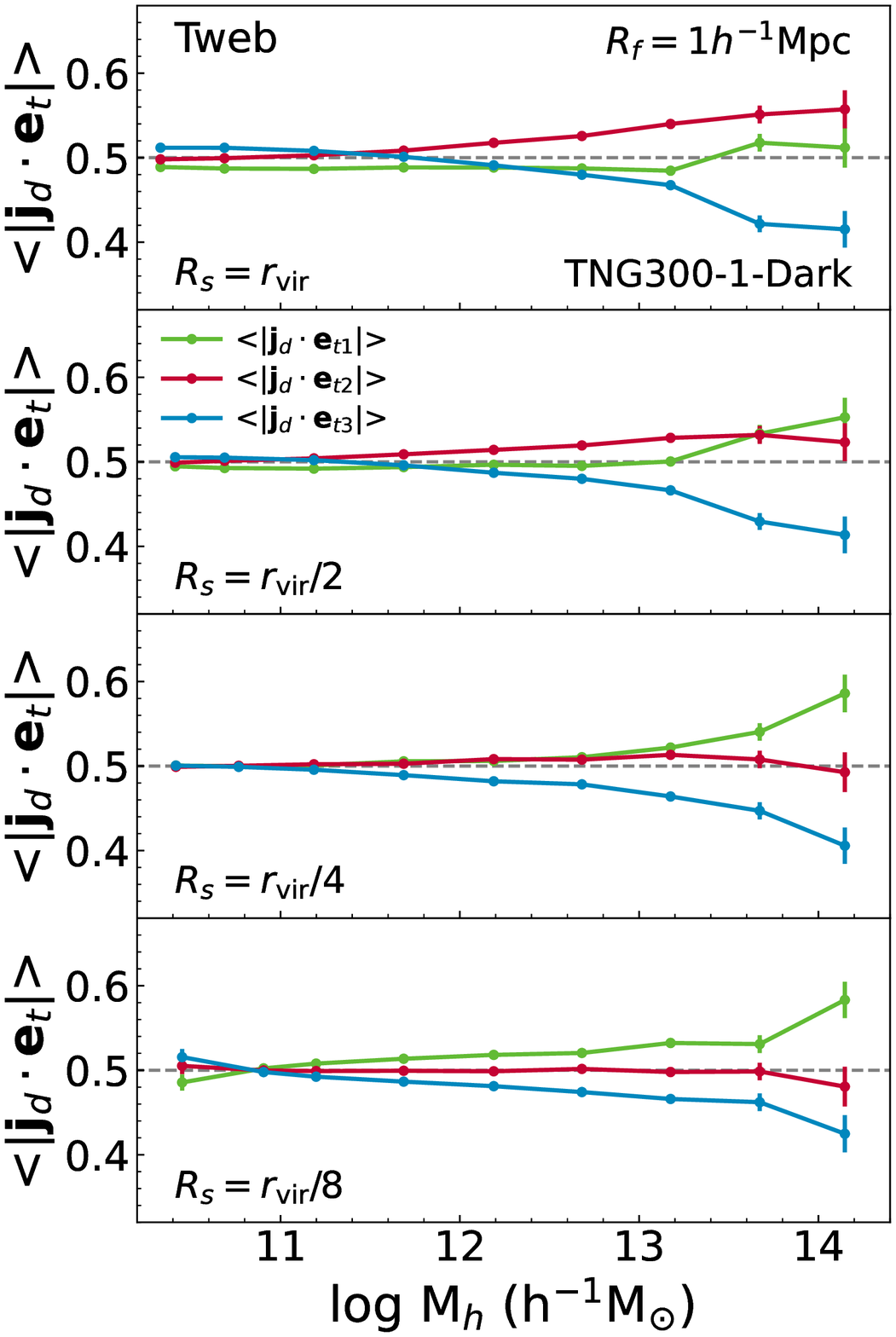}
\caption{Same as Figure \ref{fig:je_m_rf2} but for the case of $\rf=1\dunit$.}
\label{fig:je_m_rf1}
\end{figure}

\begin{figure}[ht]
\centering
\includegraphics[height=18cm,width=12cm]{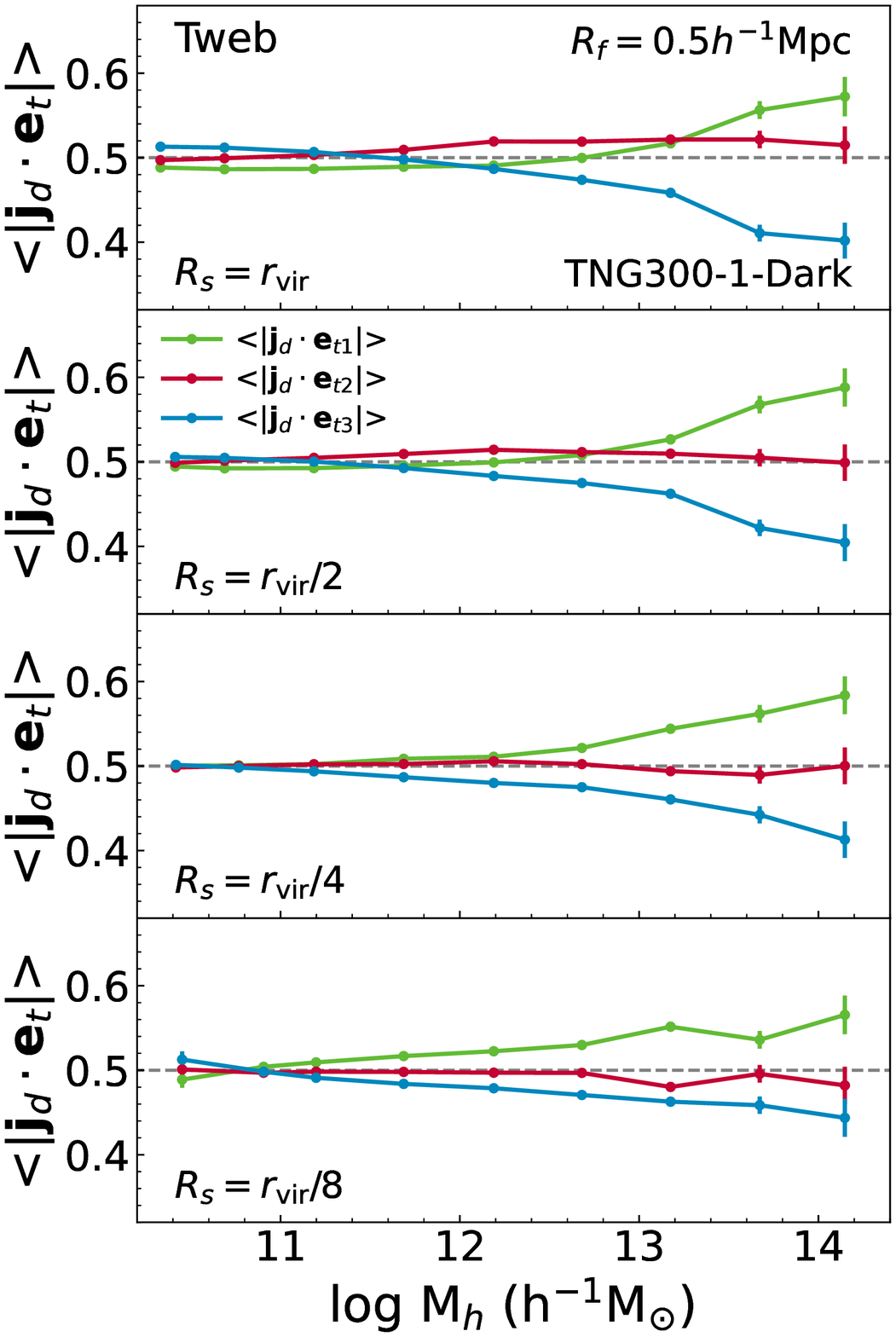}
\caption{Same as Figure \ref{fig:je_m_rf2} but for the case of $\rf=0.5\dunit$.}
\label{fig:je_m_rf0.5}
\end{figure}

Figure \ref{fig:je_m_rf1} plots the same as Figure \ref{fig:je_m_rf2} but with the Tweb smoothed on a smaller scale of $R_{f}=1\dunit$. As can be seen, 
for this case the transition of $\jd$ from $\eti$ to $\etm$ occurs at more outer radii than for the case of $R_{f}=2\dunit$. 
Even at $\rs=\rv/2$, a substantial signal of the $\jd$--$\etm$ alignment tendency is found in the highest mass bin with $M_{h}\gtrsim 10^{14}\munit$.  
Figure \ref{fig:je_m_rf0.5} plots the same as Figure \ref{fig:je_m_rf2} but on the scale of $R_{f}=0.5\dunit$. As can be seen, even the halo 
{\it virial spins} in the highest mass section with $M_{h}\gtrsim 10^{14}\munit$ appear to be aligned with the Tweb major principal axes.  
Note also that on this small scale the value of $\mth$ rapidly decreases from $10^{13}\munit$ to $10^{11}\munit$ as $\rs/\rv$ decreases from $1/2$ to 
$1/8$.  

\begin{figure}[ht]
\centering
\includegraphics[width=\textwidth]{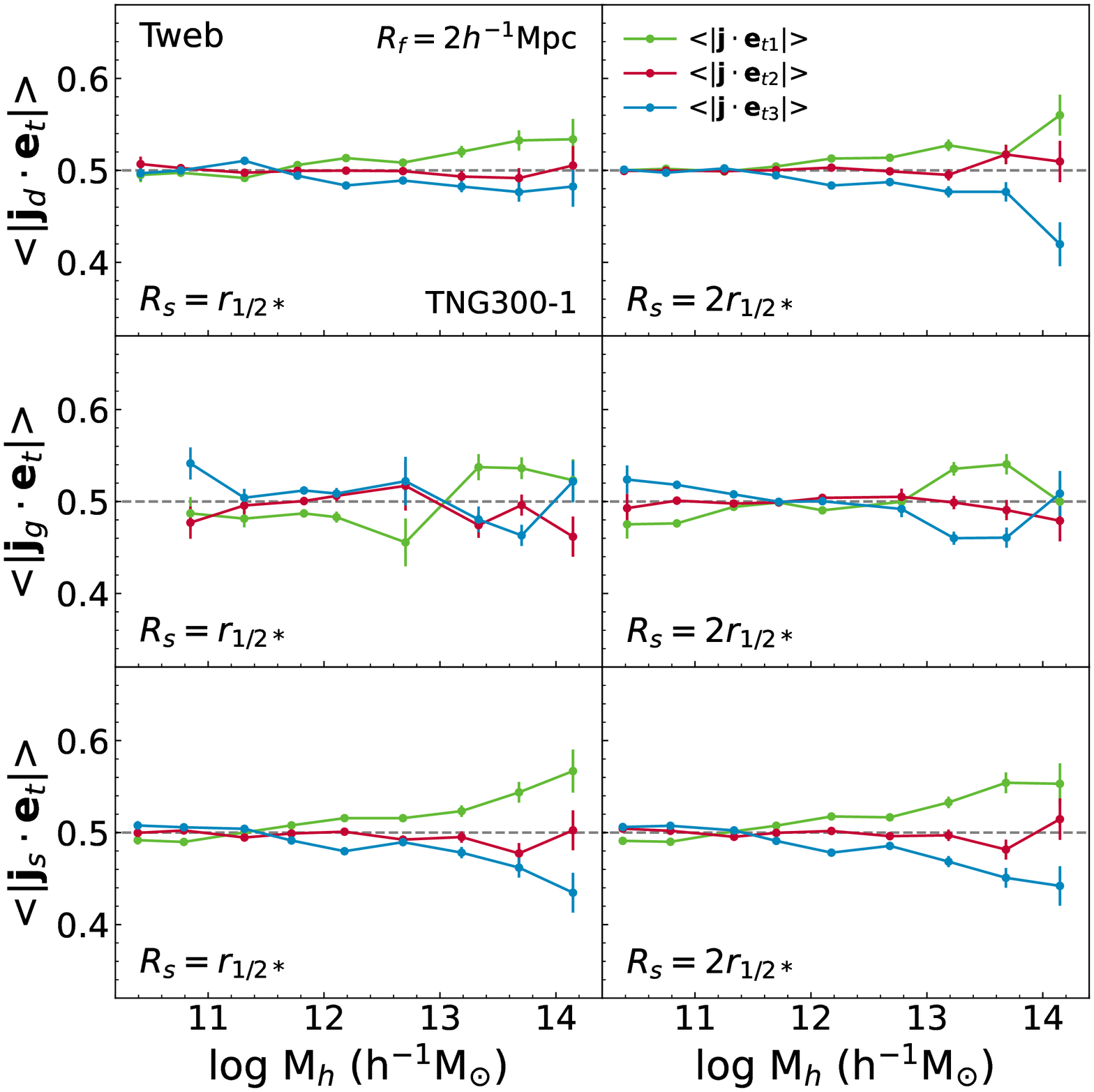}
\caption{Orientations of the spin vectors of the DM (top panels), non-stellar gas (middle panels) 
and stellar (bottom panels) components measured at $\rh$ (left panels) and $2\rh$ (right panels) in the Tweb principal frame 
on the scale of $\rf=2\dunit$ as a function of halo mass at $z=0$ from the TNG300-1 simulation.}
\label{fig:jse_m_rf2}
\end{figure}

\begin{figure}[ht]
\centering
\includegraphics[width=\textwidth]{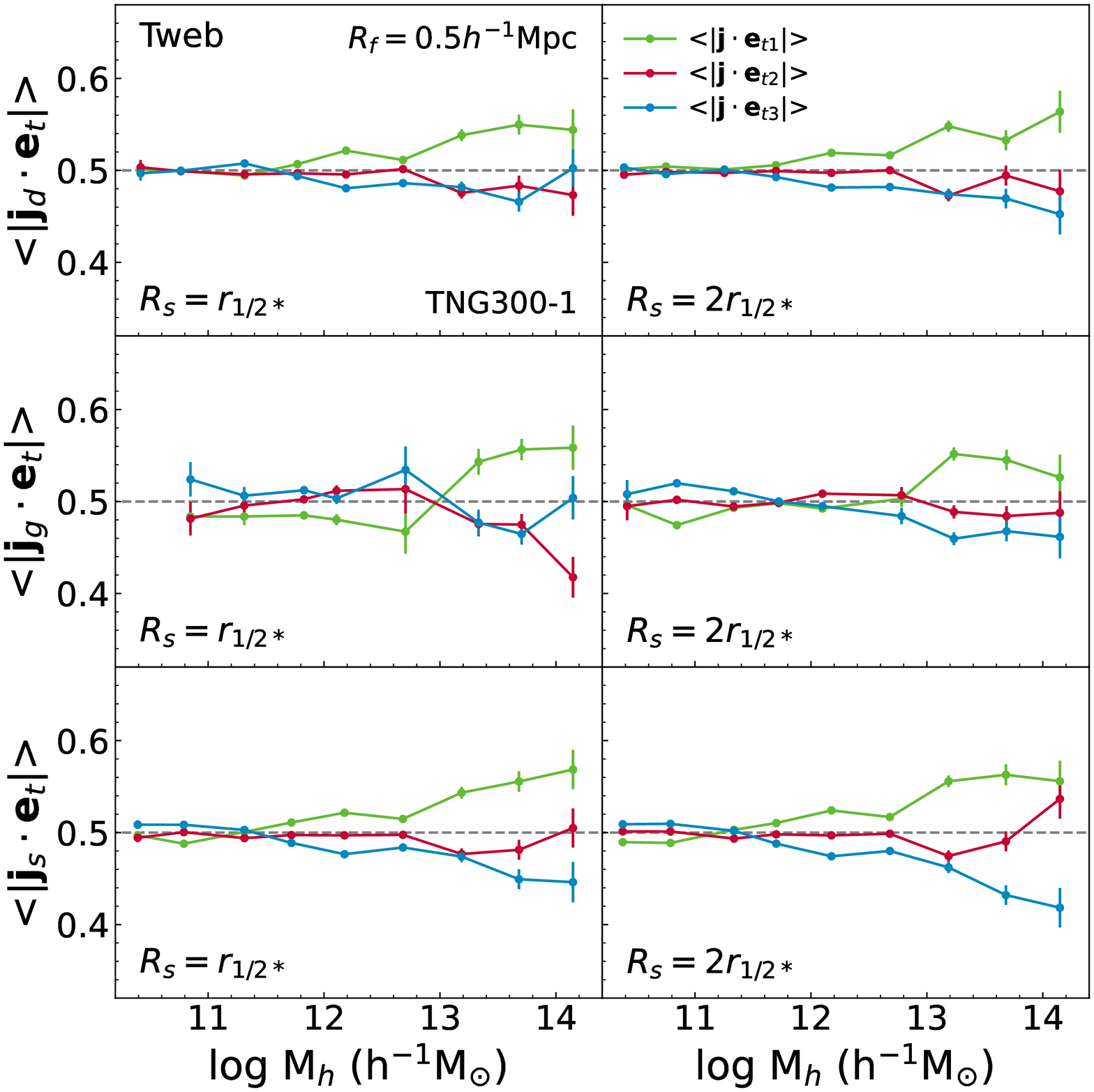}
\caption{Same as Figure \ref{fig:jse_m_rf2} but at $\rf=0.5\dunit$.}
\label{fig:jse_m_rf0.5}
\end{figure}

The halos from the TNG300-1 simulation turn out to yield similar results. Figures \ref{fig:jse_m_rf2}--\ref{fig:jse_m_rf0.5} show the same as 
Figures \ref{fig:je_m_rf2}--\ref{fig:je_m_rf0.5} but from the TNG300-1 with the inner spins of the DM (top panel), non-stellar gas (middle panel) 
and stellar components (bottom panel) measured separately at two innermost radii $\rs=2\rh$ and $\rh$. 
Note that the peculiar spin alignments with the Tweb major principal axes are found not only from the stellar components but also from the other two counterparts  
and that the $\jd$--$\etm$ and $\js$--$\etm$ alignments are more similar to each other in their tendencies than the $\jg$--$\etm$ counterpart. These results imply 
that the mechanism responsible for the peculiar spin alignment should affect not only the baryonic particles but also the DM counterparts, in contrast to 
what \citet{lee-etal21} speculated in their original work. A notable difference, however, exists between the DM and stellar components in their spin alignments 
measured at the innermost radii. 
The occurrence of the type I spin transition is witnessed from the latter but not from the former, for all the cases of $\rf$ and $\rs$. 

\begin{figure}[ht]
\centering
\includegraphics[height=18cm,width=12cm]{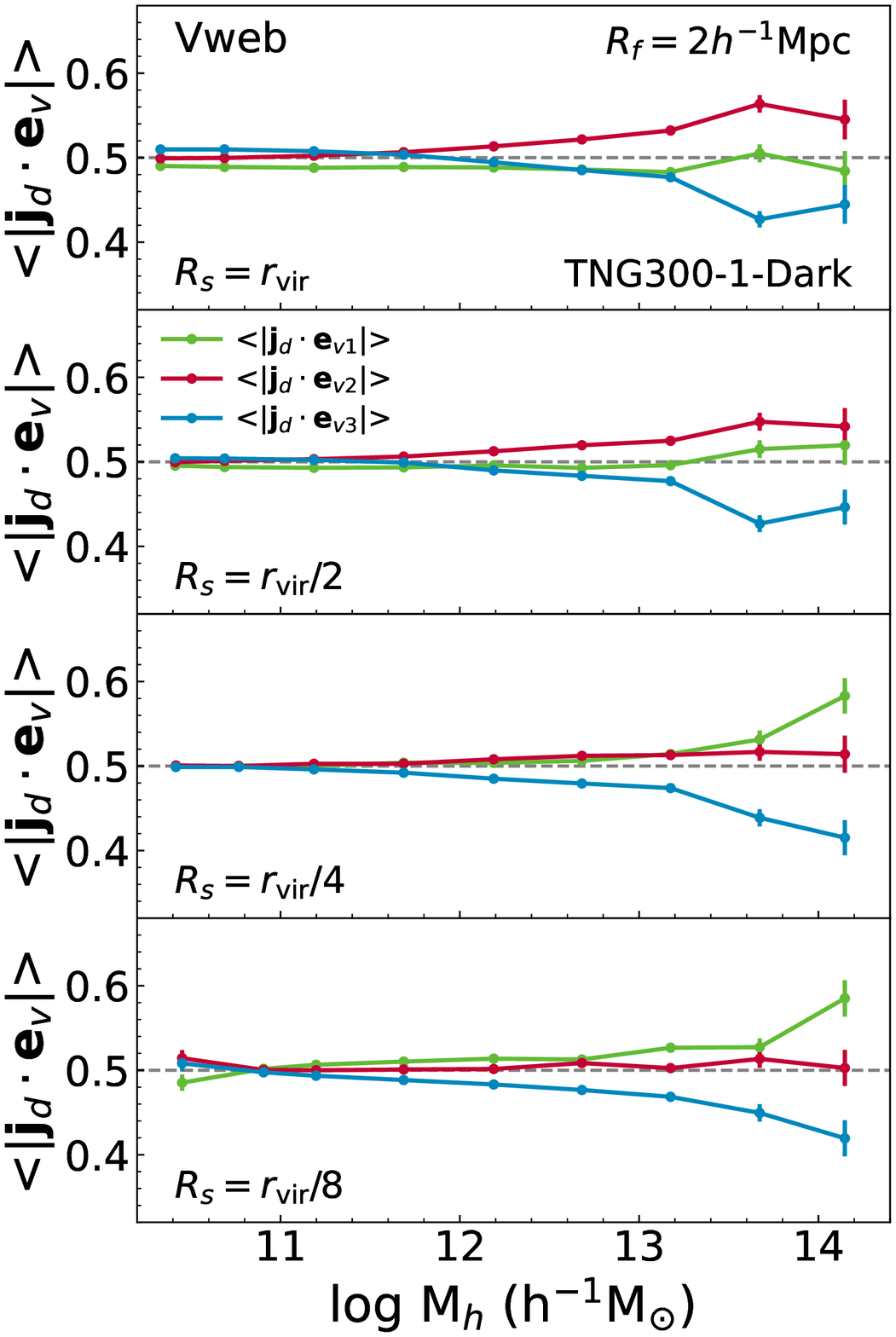}
\caption{Same as Figure \ref{fig:je_m_rf2} but with the Vweb instead of the Tweb.} 
\label{fig:jq_m_rf2}
\end{figure}

\begin{figure}[ht]
\centering
\includegraphics[height=18cm,width=12cm]{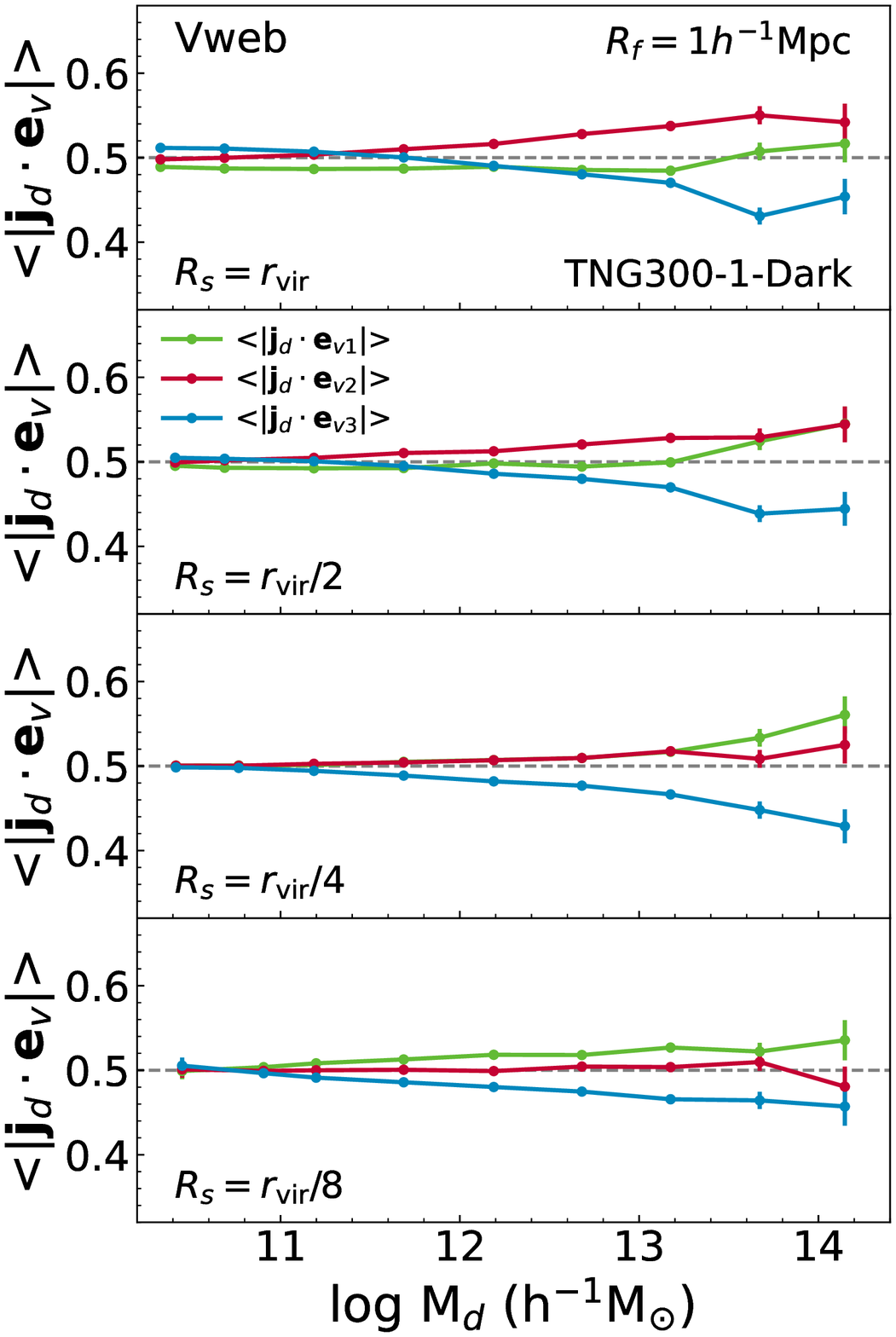}
\caption{Same as Figure \ref{fig:jq_m_rf2} for the case of $\rf=1\dunit$.}
\label{fig:jq_m_rf1}
\end{figure}

\begin{figure}[ht]
\centering
\includegraphics[height=18cm,width=12cm]{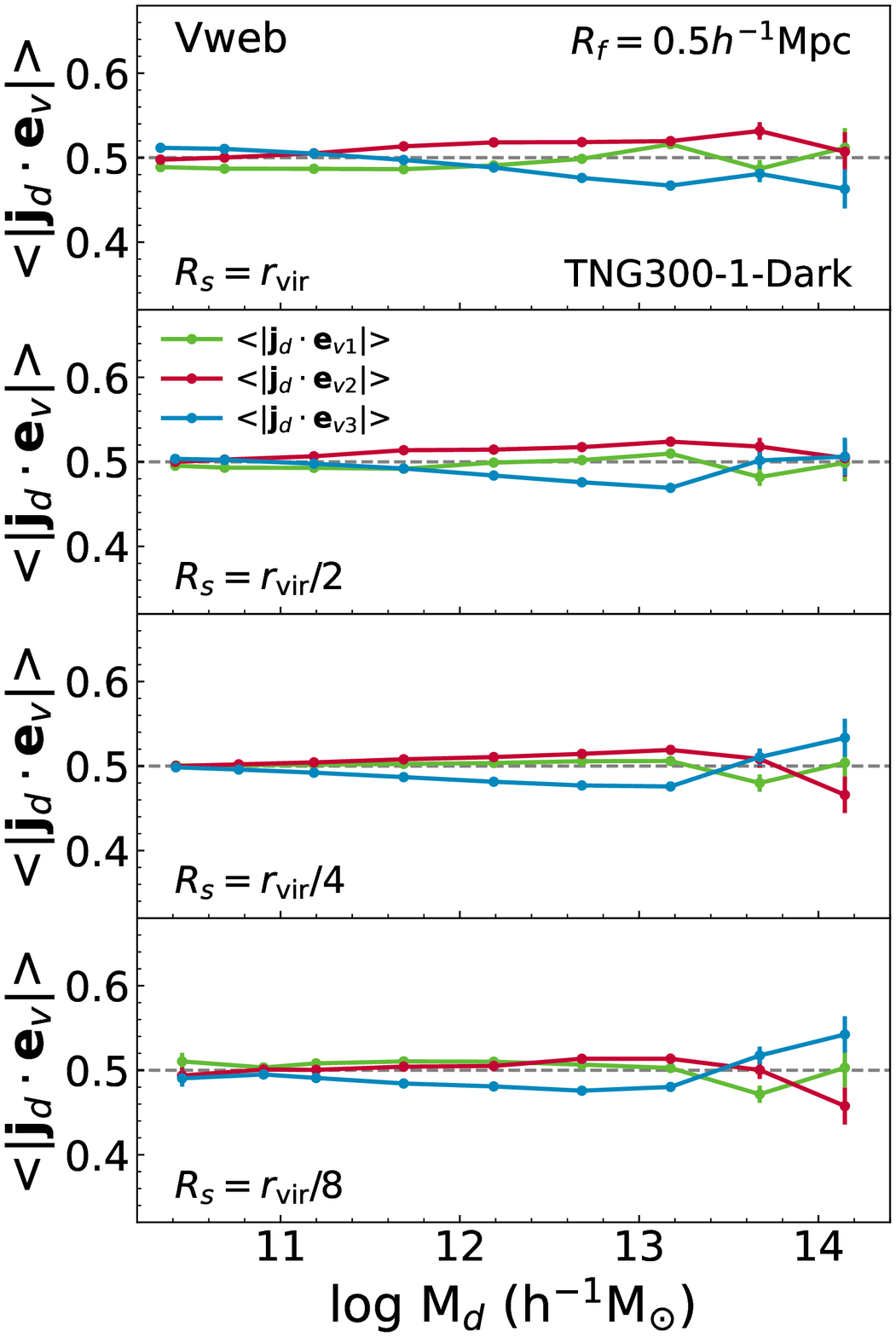}
\caption{Same as Figure \ref{fig:jq_m_rf2} for the case of $\rf=0.5\dunit$.}
\label{fig:jq_m_rf0.5}
\end{figure}

\begin{figure}[ht]
\centering
\includegraphics[width=\textwidth]{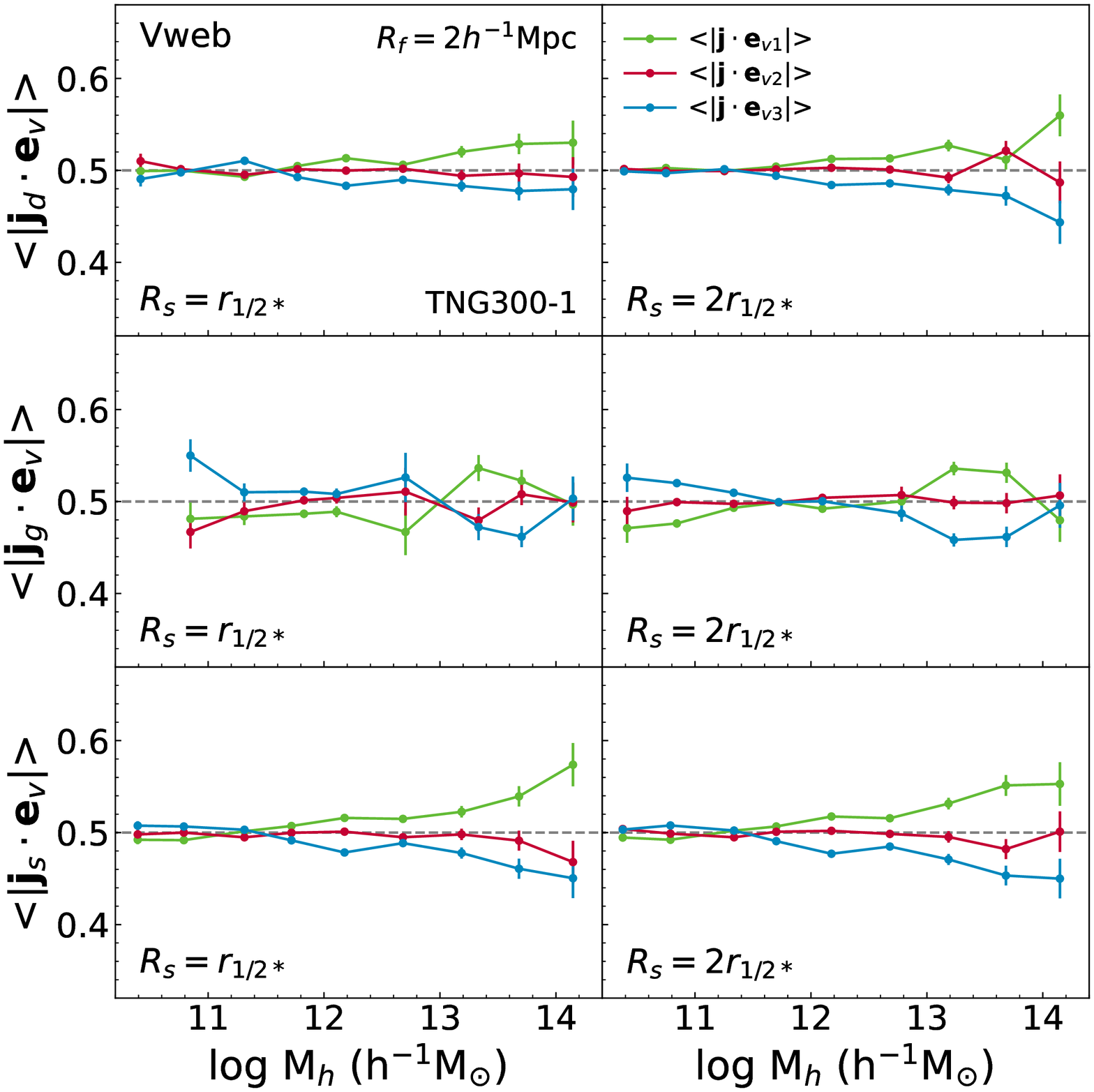}
\caption{Same as Figure \ref{fig:jse_m_rf2} but with the Vweb instead of the Tweb.} 
\label{fig:jsq_m_rf2}
\end{figure}

\begin{figure}[ht]
\centering
\includegraphics[width=\textwidth]{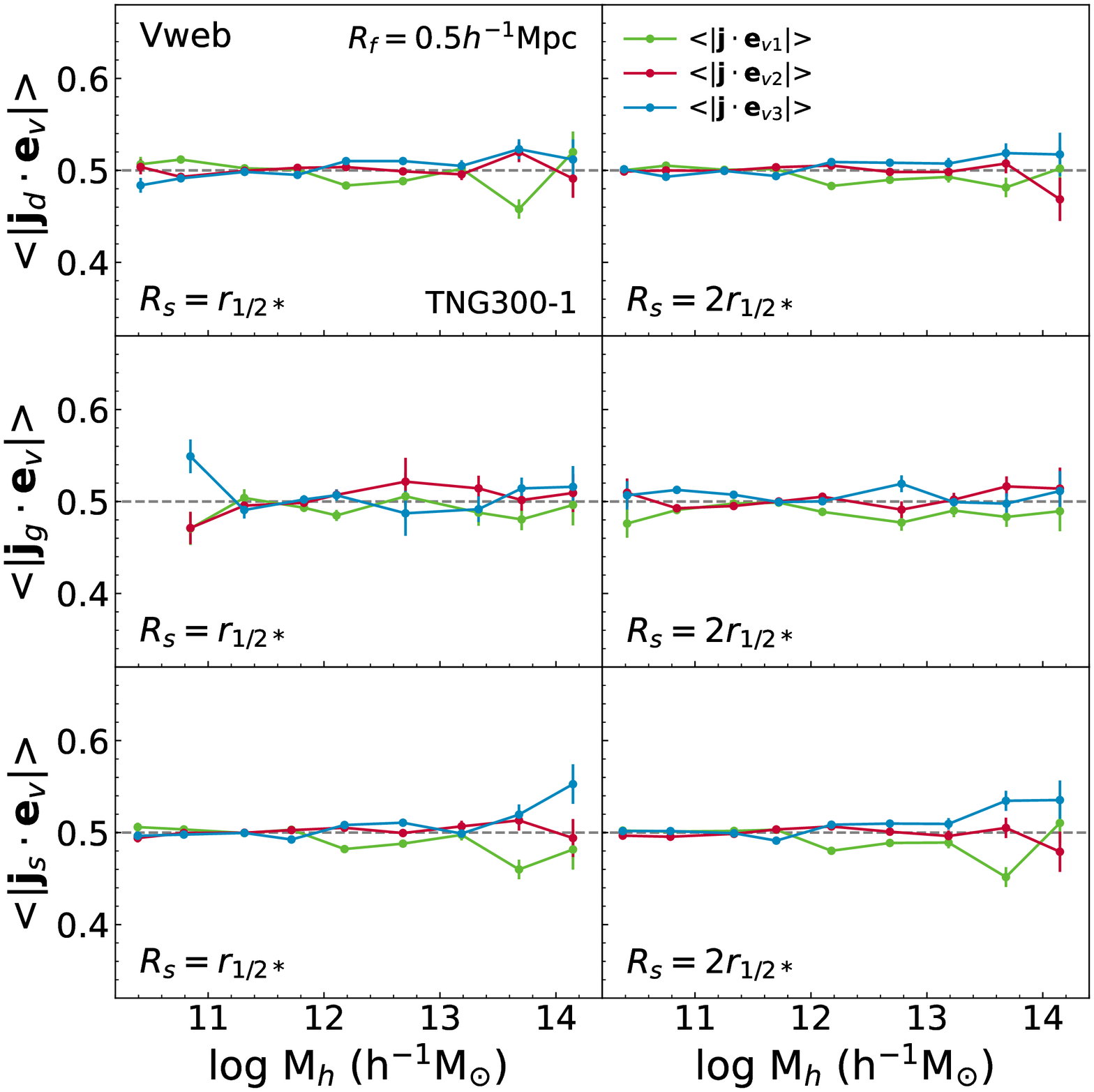}
\caption{Same as Figure \ref{fig:jsq_m_rf2} for the case of $\rf=0.5\dunit$.}
\label{fig:jsq_m_rf0.5}
\end{figure}

Figures \ref{fig:jq_m_rf2}--\ref{fig:jsq_m_rf0.5} plot the same as Figures \ref{fig:je_m_rf2}--\ref{fig:jse_m_rf0.5} but with the Vweb substituting for the Tweb. 
As can be seen, for the cases of the larger smoothing scales, $R_{f}=2\dunit$ and $1\dunit$, the results from the Vweb are very similar to those from the Tweb. 
However, a sharp difference is found for the case of the smaller smoothing scale, $R_{f}=0.5\dunit$, where no significant signal of the $\jd$--$\etm$ alignment is 
found, in direct contrast to the results from the Tweb. In fact, for this case, both of the halo inner and virial spins show no significant signal of the alignments with 
the Vweb principal axes. 
Given the well known finding that the Vweb significantly deviates from the Tweb on the nonlinear scales where the velocity field develops a strong curl mode 
\citep{lib-etal14},  one may suspect that the generation of vorticity must be related to this difference between the Tweb and the Vweb 
on the scale of $\rf=0.5\dunit$. In the following subsection, we explore the net effect of vorticity on the alignment strengths and tendencies of the halo 
spin vectors with the cosmic web.

\subsection{Vorticity Effect}\label{sec:vor}

\begin{figure}[ht]
\centering
\includegraphics[height=18cm,width=12cm]{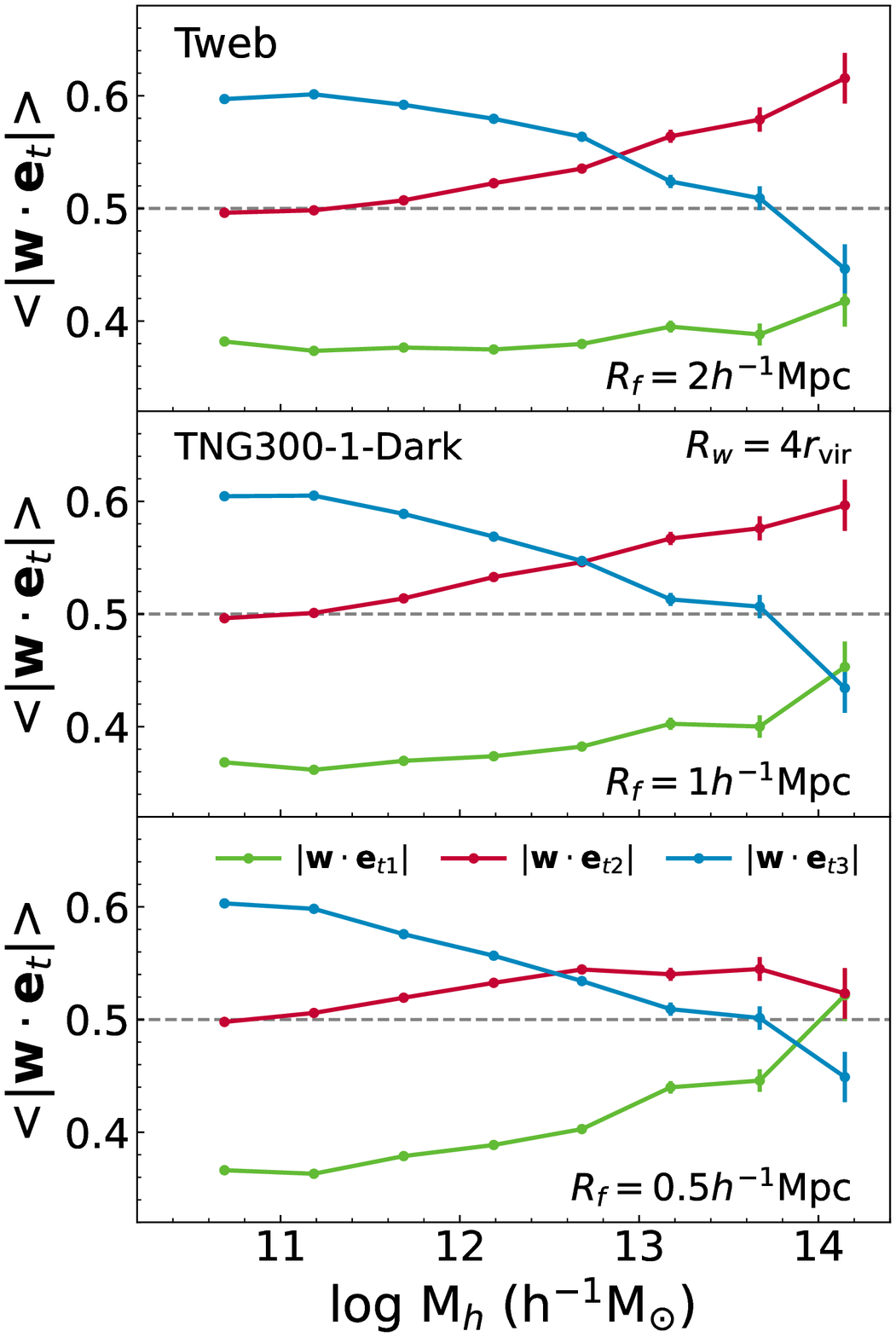}
\caption{Orientations of the vorticity vectors in the Tweb principal frame on three different scales of $\rf=2, 1$ and $0.5 \dunit$ (from top to bottom panels, respectively)
as a function of halo mass at $z=0$ from the TNG300-1-Dark simulation.}
\label{fig:vore_m}
\end{figure}

\begin{figure}[ht]
\centering
\includegraphics[height=18cm,width=12cm]{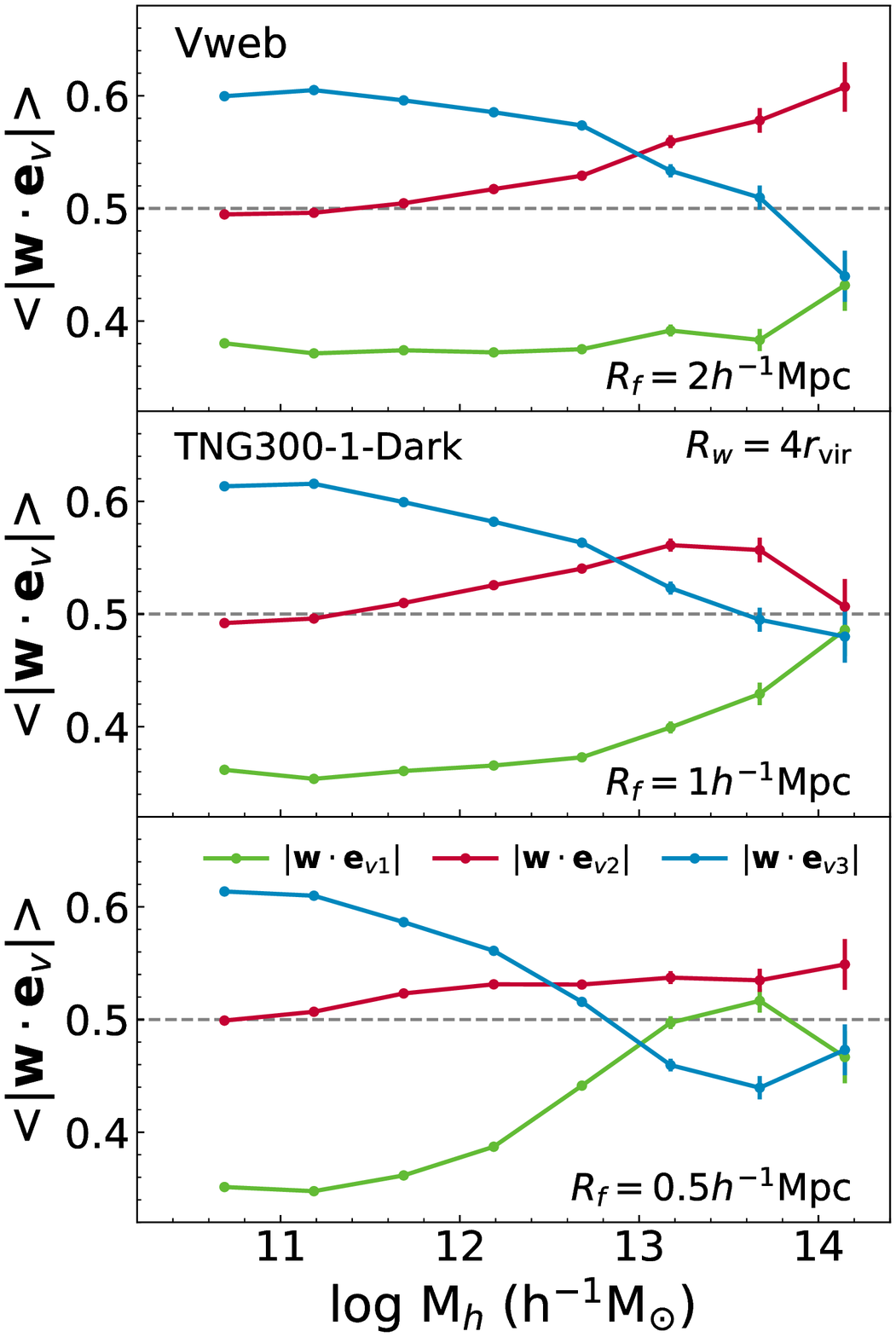}
\caption{Same as Figure \ref{fig:vore_m} but in the Vweb principal frame.}
\label{fig:vorq_m}
\end{figure}

Figures \ref{fig:vore_m}--\ref{fig:vorq_m} show how the vorticity vectors are aligned with the Tweb and Vweb principal axes, respectively, 
as a function of $M_{h}$ for the three different cases of $R_{f}$ from the TNG300-1-Dark simulation.  
As can be seen, the vorticity vectors exhibit the type II transition at some threshold mass whose value appears to increase with $R_{f}$ for both of the Tweb and 
Vweb cases, which are very similar to the spin case. 
Nevertheless, we find a marked difference in the alignment and transition tendency between the spin and vorticity orientations in 
the Tweb and Vweb principal frames. Not to mention that the overall strengths of the alignments of the vorticity vectors with the Tweb and Vweb principal axes 
are higher, no significant signal of the type I and type II transition is exhibited by the vorticity vectors even on the scale of $0.5\dunit$. Regardless of the values of 
$\rf$ and $M_{h}$, the vorticity vectors seem to be aligned with the directions perpendicular to the Tweb and Vweb major principal axes, although the alignment 
strength tends to decrease as $M_{h}$ increases. The existence of this difference implies that the halo inner spins are rather unattached to the nonlinear modification 
caused by the generation of vorticity, which the halo virial spins experience.

\begin{figure}[ht]
\centering
\includegraphics[height=18cm,width=12cm]{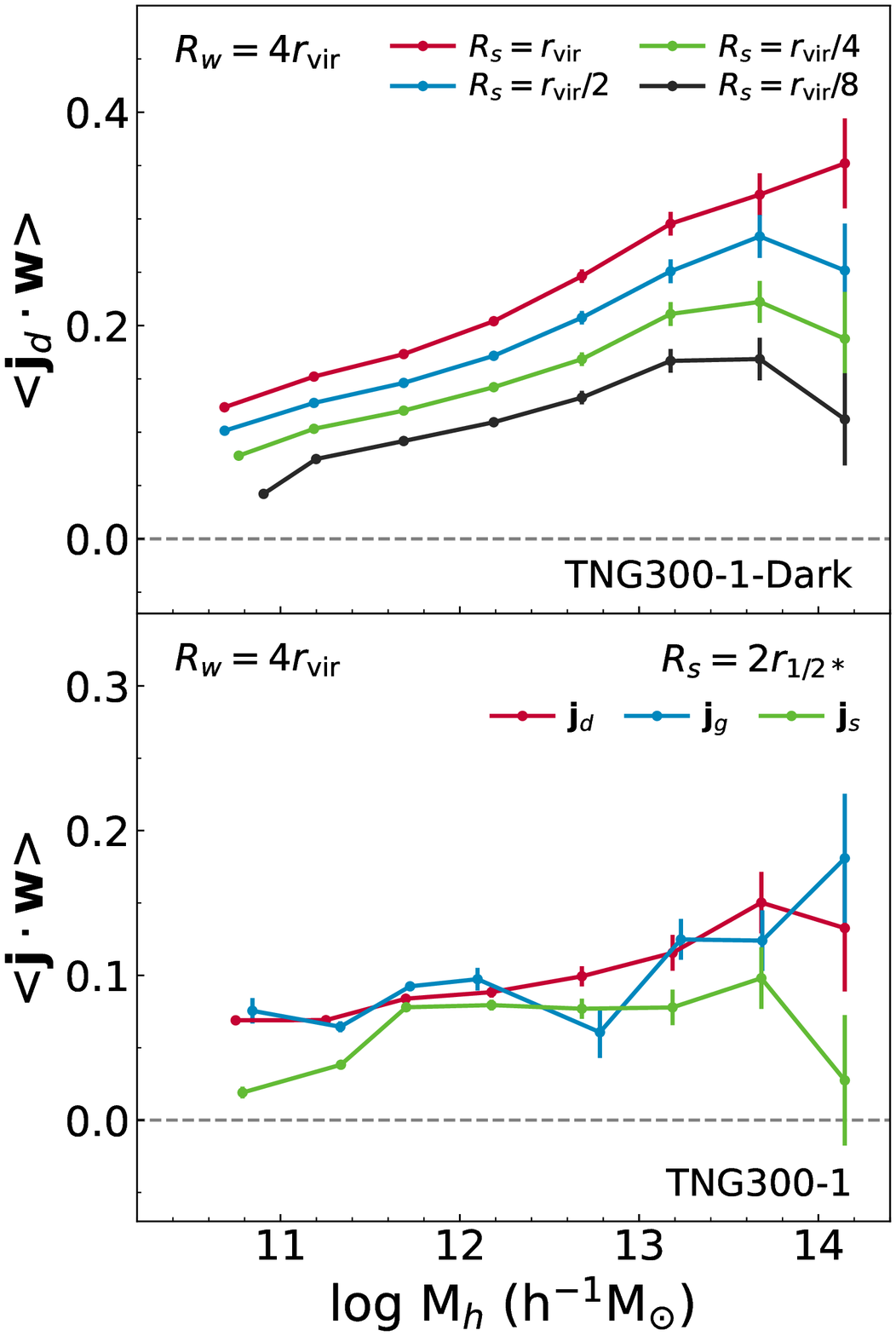}
\caption{Dot products between the halo spin and vorticity vectors as a function of the halo mass: 
halo spin vectors measured at four different radii from the TNG300-1-Dark simulation (top panel); 
spin vectors of the DM, gas, and stellar components measured at $2\rh$ from the TNG300-1 simulation
(bottom panel). }
\label{fig:jvor_m}
\end{figure}

Figure \ref{fig:jvor_m} shows how strongly the vorticity vectors are aligned with the halo virial and inner spins.  We do not take the 
absolute values of the alignment angles for this case, since the directions of the two vectors are measurable, unlike the Tweb and Vweb principal axes.
The top panel plots the $\jd$--$\vor$ alignments measured at four different radii from the TNG300-1-Dark simulation.  The halo spin vectors 
measured at $\rs=\rv$ yield a significant tendency of being aligned with the vorticity vectors in the high-mass section, consistent with the previous findings 
\citep{lib-etal13b}.  However, the strength of the $\jd$--$\vor$ alignment rapidly decreases as $M_{h}$ decreases and as $\rs$ becomes lower than $\rv$.
The bottom panel plots the $\jd$--$\vor$, $\jg$--$\vor$, and $\js$--$\vor$ alignments measured at $2\rh$ as a function of $M_{h}$ from the TNG300-1 simulation. 
As can be seen, none of the three spin vectors measured at $2\rh$ exhibit a strong alignment with the vorticity vectors, which supports the scenario 
that the halo inner spins are not susceptible to the vorticity effect. 

\begin{figure}[ht]
\centering
\includegraphics[height=14cm,width=14cm]{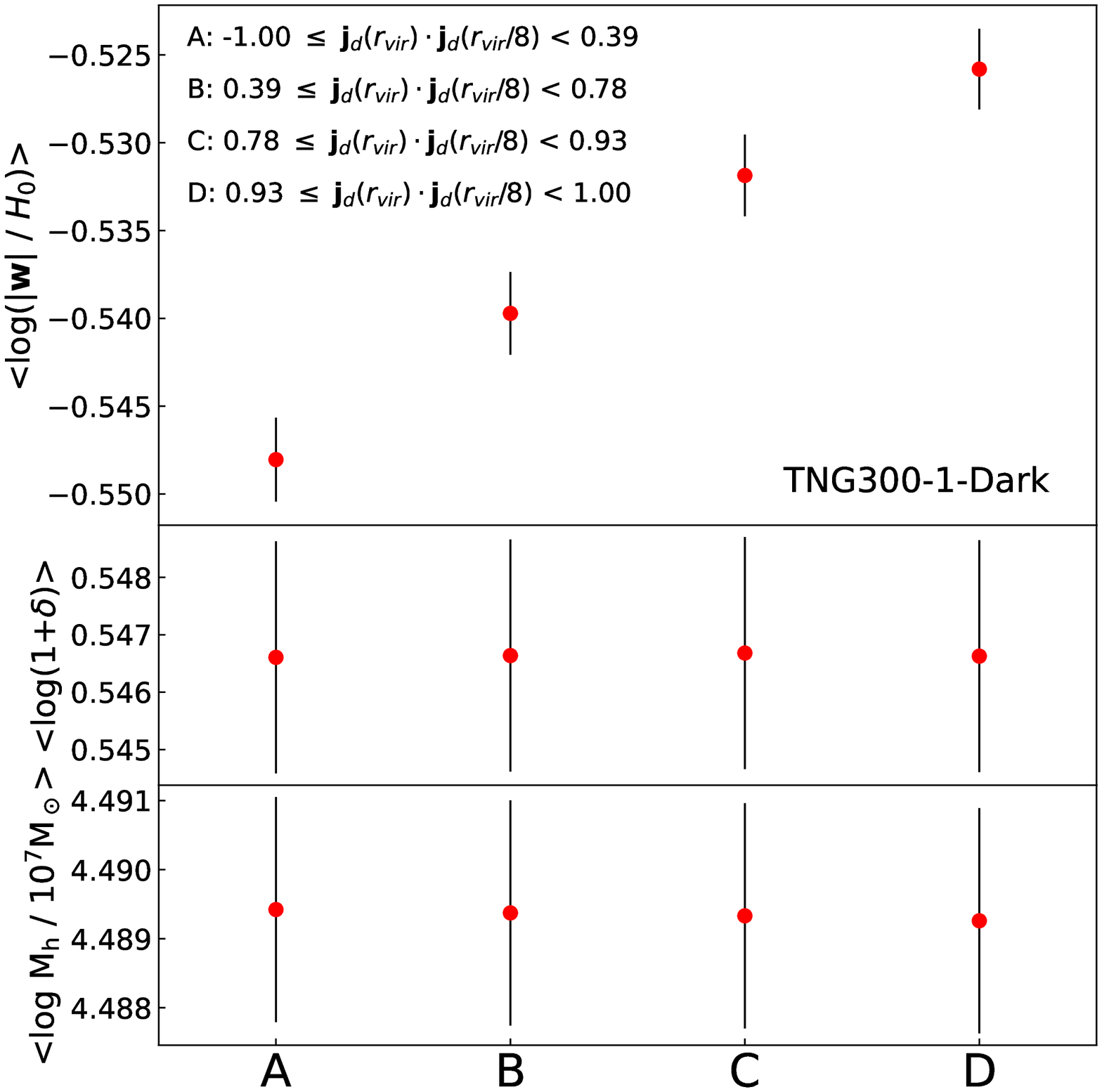}
\caption{Mean values of the logarithms of the vorticity magnitudes (top panel), densities (middle panel), and halo masses 
(bottom panel) from four controlled samples classified by the dot-products between the halo spin vectors measured at two different 
radii from the TNG300-1-Dark simulation.}
\label{fig:vor_rad}
\end{figure}

\begin{figure}[ht]
\centering
\includegraphics[height=14cm,width=14cm]{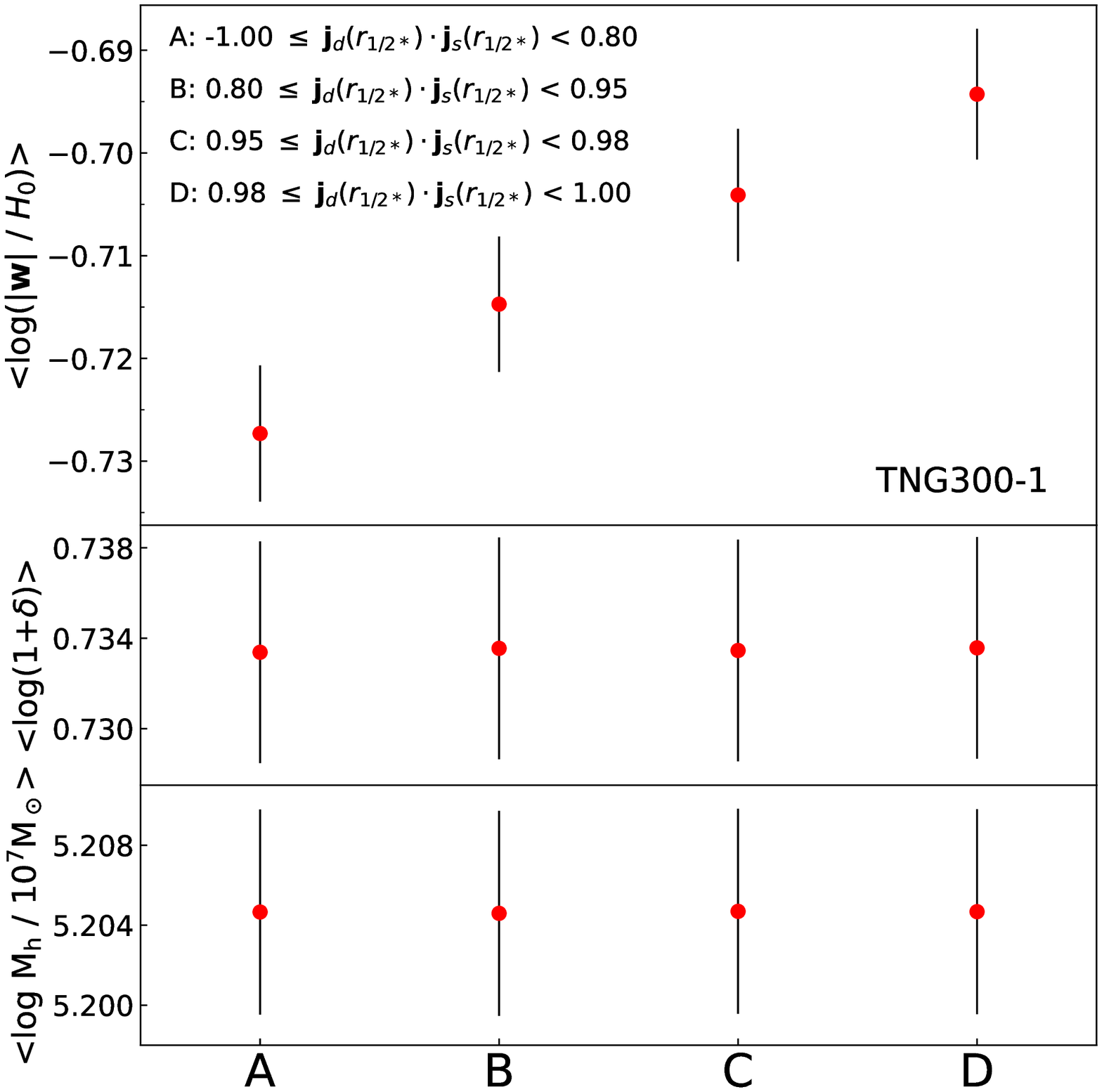}
\caption{Same as Figure \ref{fig:vor_rad} but from the controlled samples classified by the dot-products between the 
DM and stellar spin vectors from the TNG300-1 simulation.}
\label{fig:vor_dms}
\end{figure}

Figure \ref{fig:vor_rad} plots the mean magnitudes of the logarithms of the rescaled vorticity vectors, $\log\vert\vor\vert/H_{0}$ (top panel), mean density 
contrasts in the logarithmic scale (middle panel) and mean halo mass (bottom panel) averaged over each of the four controlled subsamples of the halos classified 
by the values of $\jd(\rv)\cdot\jd(\rv/8)$ from the TNG300-1-Dark simulation.
Note first that the four subsamples are indeed well controlled enough to have almost identical mean mass and density contrasts to one another. It guarantees 
that any difference in the vorticity magnitudes, if found, among the four subsamples should be attributed to the differences not in the halo mass nor in the local densities 
but in the strengths of the alignments between the spin vectors measured at the two different radii, $\rv$ and $\rv/8$. As can be seen, there is a clear signal of the 
variation of $\langle\log\vert\vor\vert/H_{0}\rangle$ with the subsamples. The more strongly the inner and virial spins are aligned with each other, the higher magnitudes 
the vorticity vectors have. 
Figure \ref{fig:vor_dms} plots the same as Figure \ref{fig:vor_rad} but with the subsamples classified by $\jd(\rh)\cdot\js(\rh)$ from the TNG300-1 simulation, 
which exhibits a similar trend. The vorticity vectors tend to have higher magnitudes when the DM and stellar spin vectors at $\rh$ are more strongly aligned 
with each other. The results shown in Figures \ref{fig:vor_rad}--\ref{fig:vor_dms} indicate that in the regions with high vortical motions the halo spin vectors 
tend to have more coherent directions across radii, regardless of the components. 

\begin{figure}[ht]
\centering
\includegraphics[height=14cm,width=14cm]{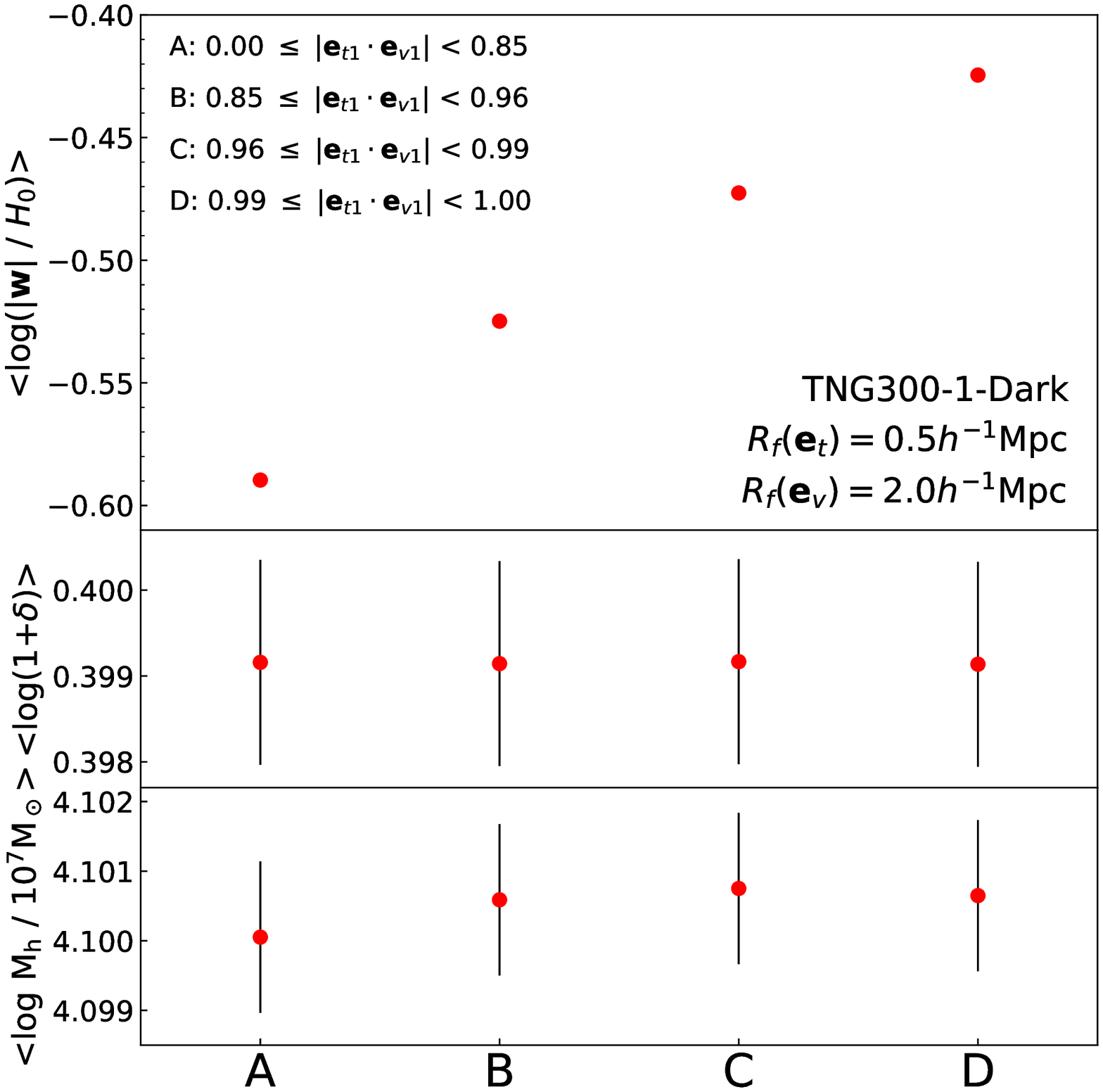}
\caption{Mean values of the logarithms of the vorticity magnitudes (top panel), densities (middle panel), and halo masses 
(bottom panel) from four controlled samples classified by the absolute values of the dot-products between the Tweb and Vweb major 
principal axes measured at two different smoothing scales from the TNG300-1-Dark simulation. The errors in the the logarithms of the vorticity 
magnitudes are invisible due to their small sizes.}
\label{fig:vor_tv_dark}
\end{figure}

\begin{figure}[ht]
\centering
\includegraphics[height=14cm,width=14cm]{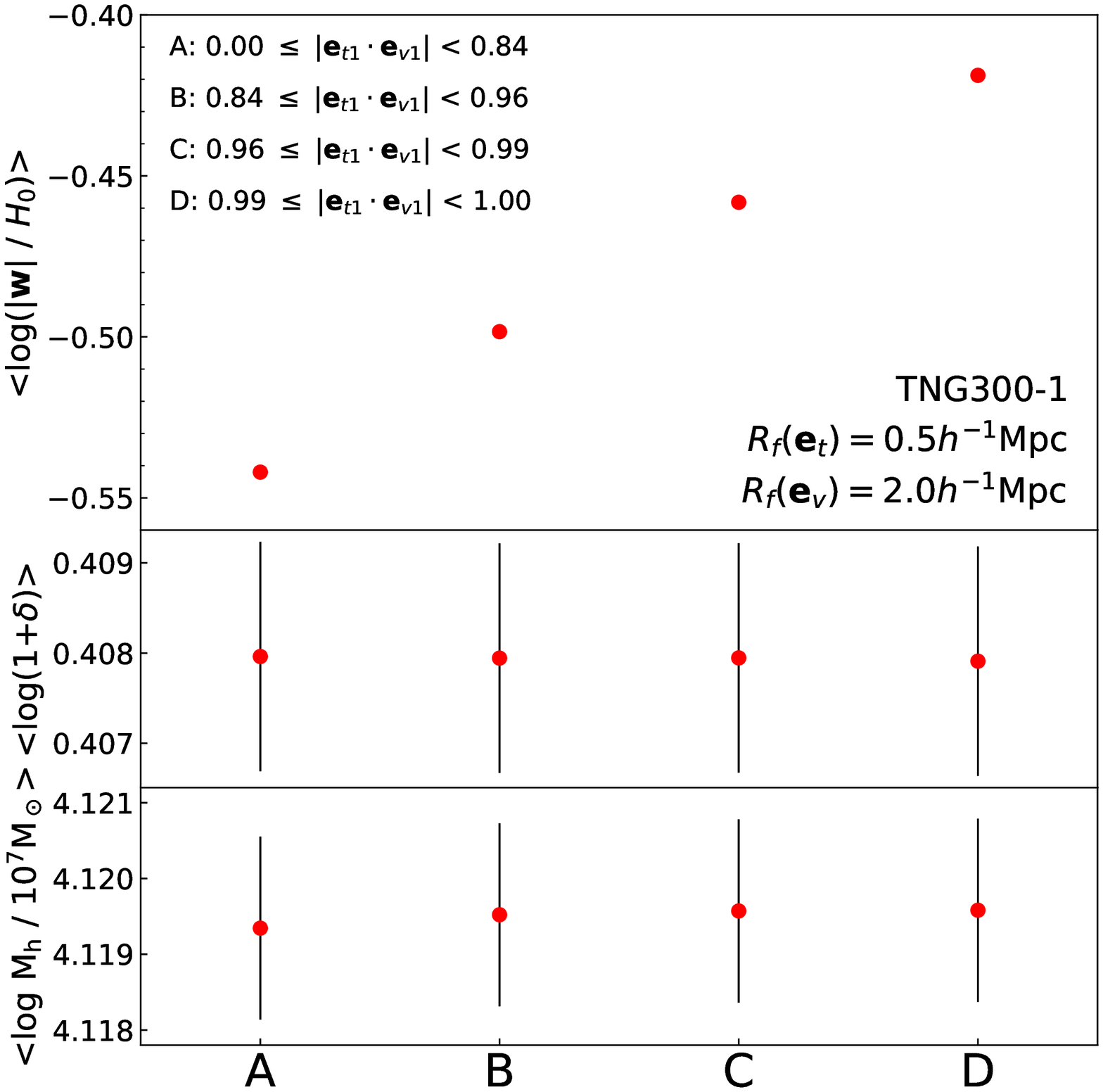}
\caption{Same as Figure \ref{fig:vor_tv_dark} from the TNG300-1 simulation.}
\label{fig:vor_tv}
\end{figure}

Figures \ref{fig:vor_tv_dark}--\ref{fig:vor_tv} plot the same as Figures \ref{fig:vor_rad}--\ref{fig:vor_dms} but with the subsamples classified by 
$\vert\etm (0.5\dunit)\cdot\evm (2\dunit)\vert$ from the TNG300-1-Dark and TNG300-1 simulations, respectively.  As can be seen, the higher vorticity magnitudes are found in the 
subsamples of the halos located in the regions where the large-scale Vweb and small-scale Tweb are more strongly aligned with each other. Note the consistency of this result with that 
shown in Figures \ref{fig:vor_rad}--\ref{fig:vor_dms}. The weaker alignments between $\etm (0.5\dunit)$ and $\evm (2\dunit)$ can be translated into the weaker alignments 
between $\jd (\rv)$ and $\jd (\rv/8)$, since the latter tends to be aligned with the major principal axes of the small-scale Tweb in the wide mass range 
while the halo virial spins are aligned with the directions perpendicular to the major principal axes of the large-scale Vweb in the whole mass range considered. 

\section{Summary and Discussion}\label{sec:sum}

We have explored how the halo spin vectors change their preferred directions with respect to the cosmic web if they are measured at inner radii smaller 
than the virial boundaries by analyzing the data from the DM-only and hydrodynamical simulations of the IllustrisTNG project 
\citep{tngintro1, tngintro2, tngintro3, tngintro4, tngintro5, illustris19}. Given that the velocity fields develop vortical motions in the nonlinear regime 
\citep{PB99,kit-etal12,lib-etal13a,hah-etal15}, we have also investigated if and how the generation of vorticity affects the alignment tendency and strength between 
the halo inner spins and the cosmic web. Using two different algorithms, called the Tweb and Vweb finders \citep{lib-etal14,for-etal14}, 
to identify the cosmic web, we have also tested their validity and efficiency in capturing the variation of the alignment tendency of the halo inner spins 
with radial distances. 

Our findings are summarized as follows:
\begin{itemize}
\item
The halos in a certain mass range from the TNG300-1-Dark simulations exhibit a radius-dependent spin transition in the principal frame of the Tweb for all of 
the three different cases of $\rf/(\dunit)=0.5,1$ and $2$. The preferred directions of their spin vectors measured at inner radii smaller than their virial boundaries, 
$\rs<\rv$,  transit from the Tweb intermediate to major principal axes where the ratio, $\rs/\rv$, is reduced down to some threshold, $\uth$.  
Meanwhile, in the Vweb principal frame, a similar radius-dependent spin transition is witnessed only for the case of $\rf\ge 1\dunit$. 
\item
The radius ratio threshold, $\uth$, sensitively depends on $\rf$,  as well as on the halo mass $M_{h}$.  For the case of $\rf\ge 1\dunit$, 
the occurrence of the radius-dependent transition is witnessed in the large mass range of $M_{h}\ge 10^{12}\munit$. The more massive the halos are, the larger value 
$\uth$ has.  For the case $\rf=0.5\dunit$,  however, the massive halos with $M_{h}\gtrsim 10^{13}\munit$ show alignments with the Tweb {\it major} principal axes even 
at $\uth=1$, which implies that the alignments between the halo inner spins and the Tweb major principal axes occur in the mass bin with corresponding inner radii 
greater than $\rf$. 
\item
The inner spins of the halo DM components measured at $\rh$ and $2\rh$ from the TNG300-1 simulations exhibit the same peculiar alignments with the 
Tweb major principal axes as those of the halo non-stellar gas and stellar components. However, the occurrence of the mass-dependent type I spin transition 
between the Tweb major and minor principal axes is witnessed only from the gas and stellar spin vectors but not from the DM counterparts. 
\item
Unlike the halo spins, the vorticity vectors show no significant tendency of being aligned with the Tweb major principal axes, regardless 
of $M_{h}$ and $\rf$. But, it has a net effect of enhancing the alignments between the halo inner and virial spins, 
the alignments between the DM and stellar spins measured at $\rh$, and the alignments between the major principal axes of the small-scale 
Tweb and large-scale Vweb. 
\end{itemize}

Three key implications of our results are the following.  First, although the vorticity developed in the nonlinear regime may originate the halo virial spins as claimed 
by \citet{lib-etal13b}, it must have little effect on the directions of the halo inner spins given the obvious difference in their alignment tendencies with the Tweb 
major principal axes. Second, unlike what \citet{lee-etal21} originally speculated, the peculiar alignment tendency of the halo stellar spins 
at $2\rh$ with the Tweb major principal axes should not be ascribed to any baryonic effects.  Rather, the alignments of the stellar 
vectors at $2\rh$ with the Tweb {\it minor} principal axes are likely to be enhanced by some non-gravitational baryonic effects, given that the DM spins 
at $2\rh$ show no alignments with the Tweb minor principal axes even in the lowest mass section.  Some baryonic process like galactic winds that can discharge 
stellar materials from the low-mass halos might occur anisotropically along the directions of filaments, which eventually lead the stellar spins at $2\rh$ to acquire a tendency 
of being aligned with the Tweb minor principal axes \citep[e.g.,][]{ten-etal17}. 
Third, on the small scale of $\rf\simeq 0.5\dunit$, the Tweb principal frame is more efficient in capturing the peculiar alignments of the halo inner spins than the Vweb counterpart. 

We speculate that the peculiar alignments of the halo inner spins with the Tweb major principal axes may be a fossil record of the gravitational process that 
the halo progenitors undergo at early epochs when the velocity field has yet to develop full vortical flows. Recall the result of \citet{LM22} that the stellar spin 
vectors at $2\rh$ are more strongly aligned with the Tweb major principal axes if the halos experience the latest merger events at earlier times.  
During the quiescent time without experiencing any merger, the interior of a halo can retain the tendency of its spin being aligned with the Tweb major principal 
axes. The merging events along the nonlinear cosmic web may stir up the interior particles of halos and drive them to have spins aligned with the directions 
perpendicular to the Tweb major principal axes, creating coherence in the spin directions across inner radii. 

Recall also what \citet{ver-etal11} discovered by analyzing the Milky-way size galactic halos from a high-resolution simulation about the halo shape profiles. 
The halo shapes measured at different radii turned out to reflect well their progenitor history, gradually switching from being prolate to being oblate as the radial distances 
increase from the innermost regions out to the virial boundaries. \citet{ver-etal11} explained that the differences in the merging directions along the cosmic web between 
the past and present epochs must be the origin of this radius-dependent shape transitions of galactic halos. 
We claim here that our results can also be explained by a similar logic.  When the matter becomes compressed first along the Tweb major principal axes at early epochs 
where no vortical motions are yet to be generated, it creates peculiar spin alignments. In the subsequent evolutions, when the second and third collapse proceeds 
at later epochs where the velocity field begins to develop a curl mode, the halo virial spins acquire an opposite tendency of being perpendicular to the Tweb major 
principal axes, while the inner spins still retain the memory of the earlier merging history. 

We believe that the halo spin alignment profile contains more information on the history of merging events along the cosmic web than the halo shape profile since the 
former is determined not only by the spatial distributions of particles but also by their velocities unlike the latter. To concrete this scenario and to test it against real 
observations, however, a much more comprehensive numerical analysis must be done by tracking down the directions of infalling DM materials at all different epochs, 
which is beyond the scope of this paper. We intend to pursue a follow-up work in this direction, hoping to report the results elsewhere in the near future. 

\acknowledgments

We thank an anonymous referee for useful comments which helped us improve the original manuscript. 
The IllustrisTNG simulations were undertaken with compute time awarded by the Gauss Centre for Supercomputing (GCS) 
under GCS Large-Scale Projects GCS-ILLU and GCS-DWAR on the GCS share of the supercomputer Hazel Hen at the High 
Performance Computing Center Stuttgart (HLRS), as well as on the machines of the Max Planck Computing and Data Facility 
(MPCDF) in Garching, Germany. 
JSM acknowledges the support by the NRF of Korea grant funded by the Korean government (MEST) (No. 2019R1A6A1A10073437. 
JL acknowledges the support by Basic Science Research Program through the National Research Foundation (NRF) of Korea 
funded by the Ministry of Education (No.2019R1A2C1083855).

\clearpage
\clearpage
\clearpage
\clearpage
\clearpage
\clearpage
\clearpage
\clearpage
\clearpage
\clearpage
\clearpage
\clearpage
\clearpage
\clearpage
\clearpage
\clearpage

\clearpage
\begin{thebibliography}{000}
\bibitem[Arag{\'o}n-Calvo et al.(2007)]{ara-etal07} 
Arag{\'o}n-Calvo, M.~A., van de Weygaert, R., Jones, B.~J.~T., et al.\ 2007, \apjl, 655, L5.
\bibitem[Aragon-Calvo \& Yang(2014)]{AY14} 
Aragon-Calvo, M.~A., \& Yang, L.~F.\ 2014, \mnras, 440, L46
\bibitem[Bett et al.(2007)]{bet-etal07} 
Bett, P., Eke, V., Frenk, C.~S., et al.\ 2007, \mnras, 376, 21
\bibitem[Coccato et al.(2009)]{coc-etal09} 
Coccato, L., Gerhard, O., Arnaboldi, M., et al.\ 2009, \mnras, 394, 1249
\bibitem[Codis et al.(2012)]{cod-etal12} 
Codis, S., Pichon, C., Devriendt, J., et al.\ 2012, \mnras, 427, 3320
\bibitem[Codis et al.(2015a)]{cod-etal15a} 
Codis, S., Gavazzi, R., Dubois, Y., et al.\ 2015, \mnras, 448, 3391
\bibitem[Codis et al.(2015b)]{cod-etal15b} 
Codis, S., Pichon, C., \& Pogosyan, D.\ 2015, \mnras, 452, 3369
\bibitem[Codis et al.(2018)]{cod-etal18} 
Codis, S., Jindal, A., Chisari, N.~E., et al.\ 2018, \mnras, 481, 4753
\bibitem[Cortese et al.(2016)]{cor-etal16} 
Cortese, L., Fogarty, L.~M.~R., Bekki, K., et al.\ 2016, \mnras, 463, 170
\bibitem[Dubois et al.(2014)]{dub-etal14} 
Dubois, Y., Pichon, C., Welker, C., et al.\ 2014, \mnras, 444, 1453
\bibitem[Emsellem et al.(2007)]{ems-etal07} 
Emsellem, E., Cappellari, M., Krajnovi{\'c}, D., et al.\ 2007, \mnras, 379, 401
\bibitem[Forero-Romero et al.(2014)]{for-etal14} 
Forero-Romero, J.~E., Contreras, S., \& Padilla, N.\ 2014, \mnras, 443, 1090
\bibitem[Ganeshaiah Veena et al.(2018)]{gan-etal18} 
Ganeshaiah Veena, P., Cautun, M., van de Weygaert, R., et al.\ 2018, \mnras, 481, 414
\bibitem[Ganeshaiah Veena et al.(2019)]{gan-etal19} 
Ganeshaiah Veena, P., Cautun, M., Tempel, E., et al.\ 2019, \mnras, 487, 1607
\bibitem[Gonz{\'a}lez et al.(2017)]{gon-etal17} 
Gonz{\'a}lez, R.~E., Prieto, J., Padilla, N., et al.\ 2017, \mnras, 464, 4666
\bibitem[Hahn et al.(2007)]{hah-etal07} 
Hahn, O., Carollo, C.~M., Porciani, C., et al.\ 2007, \mnras, 381, 41
\bibitem[Hahn et al.(2010)]{hah-etal10} 
Hahn, O., Teyssier, R., \& Carollo, C.~M.\ 2010, \mnras, 405, 274
\bibitem[Hahn et al.(2015)]{hah-etal15} 
Hahn, O., Angulo, R.~E., \& Abel, T.\ 2015, \mnras, 454, 3920.
\bibitem[Hirv et al.(2017)]{hir-etal17} 
Hirv, A., Pelt, J., Saar, E., et al.\ 2017, \aap, 599, A31
\bibitem[Joachimi et al.(2015)]{align_review1} 
Joachimi, B., Cacciato, M., Kitching, T.~D., et al.\ 2015, \ssr, 193, 1.
\bibitem[Kiessling et al.(2015)]{align_review2} 
Kiessling, A., Cacciato, M., Joachimi, B., et al.\ 2015, \ssr, 193, 67.
\bibitem[Kitaura et al.(2012)]{kit-etal12} 
Kitaura, F.-S., Angulo, R.~E., Hoffman, Y., et al.\ 2012, \mnras, 425, 2422. 
\bibitem[Kraljic et al.(2020)]{kra-etal20} 
Kraljic, K., Dav{\'e}, R., \& Pichon, C.\ 2020, \mnras, 493, 362
\bibitem[Krolewski et al.(2019)]{kro-etal19} 
Krolewski, A., Ho, S., Chen, Y.-C., et al.\ 2019, \apj, 876, 52
\bibitem[Laigle et al.(2015)]{lai-etal15} 
Laigle, C., Pichon, C., Codis, S., et al.\ 2015, \mnras, 446, 2744
\bibitem[Lee \& Pen(2000)]{lp00} 
Lee, J. \& Pen, U.-L.\ 2000, \apjl, 532, L5. 
\bibitem[Lee \& Pen(2001)]{lp01} 
Lee, J. \& Pen, U.-L.\ 2001, \apj, 555, 106. 
\bibitem[Lee \& Erdogdu(2007)]{LE07} 
Lee, J. \& Erdogdu, P.\ 2007, \apj, 671, 1248
\bibitem[Lee et al.(2020)]{lee-etal20} 
Lee, J., Libeskind, N.~I., \& Ryu, S.\ 2020, \apjl, 898, L27
\bibitem[Lee et al.(2021)]{lee-etal21} 
Lee, J., Moon, J.-S., Ryu, S., et al.\ 2021, \apj, 922, 6.
\bibitem[Lee et al.(2022)]{lee-etal22} 
Lee, J., Moon, J.-S., \& Yoon, S.-J.\ 2022, \apj, 927, 29. 
\bibitem[Lee \& Moon(2022)]{LM22} 
Lee, J. \& Moon, J.-S.\ 2022, \apj, 936, 119.
\bibitem[Libeskind et al.(2013a)]{lib-etal13a} 
Libeskind, N.~I., Hoffman, Y., Forero-Romero, J., et al.\ 2013, \mnras, 428, 2489
\bibitem[Libeskind et al.(2013b)]{lib-etal13b} 
Libeskind, N.~I., Hoffman, Y., Steinmetz, M., et al.\ 2013, \apjl, 766, L15
\bibitem[Libeskind et al.(2014)]{lib-etal14} 
Libeskind, N.~I., Hoffman, Y., \& Gottl{\"o}ber, S.\ 2014, \mnras, 441, 1974
\bibitem[Marinacci et al.(2018)]{tngintro1} 
Marinacci, F., Vogelsberger, M., Pakmor, R., et al.\ 2018, \mnras, 480, 5113
\bibitem[Naiman et al.(2018)]{tngintro2} 
Naiman, J.~P., Pillepich, A., Springel, V., et al.\ 2018, \mnras, 477, 1206
\bibitem[Navarro, Frenk \& White(1996)]{nfw96} 
Navarro, J.~F., Frenk, C.~S., \& White, S.~D.~M.\ 1996, \apj, 462, 563
\bibitem[Nelson et al.(2018)]{tngintro3} 
Nelson, D., Pillepich, A., Springel, V., et al.\ 2018, \mnras, 475, 624
\bibitem[Nelson et al.(2019)]{illustris19} 
Nelson, D., Springel, V., Pillepich, A., et al.\ 2019, Computational Astrophysics and Cosmology, 6, 2 
\bibitem[Paz et al.(2008)]{paz-etal08} 
Paz, D.~J., Stasyszyn, F., \& Padilla, N.~D.\ 2008, \mnras, 389, 1127
\bibitem[Pichon \& Bernardeau(1999)]{PB99} 
Pichon, C. \& Bernardeau, F.\ 1999, \aap, 343, 663
\bibitem[Pichon et al.(2011)]{pic-etal11} 
Pichon, C., Pogosyan, D., Kimm, T., et al.\ 2011, \mnras, 418, 2493
\bibitem[Pillepich et al.(2018)]{tngintro4} 
Pillepich, A., Nelson, D., Hernquist, L., et al.\ 2018, \mnras, 475, 648
\bibitem[Planck Collaboration et al.(2016)]{planck16} 
Planck Collaboration, Ade, P.~A.~R., Aghanim, N., et al.\ 2016, \aap, 594, A1
\bibitem[Pueblas \& Scoccimarro(2009)]{PS09} Pueblas, S. \& Scoccimarro, R.\ 2009, \prd, 80, 043504
\bibitem[Romanowsky et al.(2003)]{rom-etal03} 
Romanowsky, A.~J., Douglas, N.~G., Arnaboldi, M., et al.\ 2003, Science, 301, 1696
\bibitem[Shi et al.(2015)]{shi-etal15} 
Shi, J., Wang, H., \& Mo, H.~J.\ 2015, \apj, 807, 37
\bibitem[Springel et al.(2001)]{subfind} 
Springel, V., White, S.~D.~M., Tormen, G., et al.\ 2001, \mnras, 328, 726
\bibitem[Springel et al.(2018)]{tngintro5} 
Springel, V., Pakmor, R., Pillepich, A., et al.\ 2018, \mnras, 475, 676
\bibitem[Springel(2010)]{arepo} 
Springel, V.\ 2010, \araa, 48, 391. doi:10.1146/annurev-astro-081309-130914
\bibitem[Tempel \& Libeskind(2013)]{TL13} 
Tempel, E., \& Libeskind, N.~I.\ 2013, \apjl, 775, L42
\bibitem[Tempel et al.(2013)]{tem-etal13} 
Tempel, E., Stoica, R.~S., \& Saar, E.\ 2013, \mnras, 428, 1827
\bibitem[Tenneti et al.(2017)]{ten-etal17} 
Tenneti, A., Gnedin, N.~Y., \& Feng, Y.\ 2017, \apj, 834, 169
\bibitem[Trowland et al.(2013)]{tro-etal13} 
Trowland, H.~E., Lewis, G.~F., \& Bland-Hawthorn, J.\ 2013, \apj, 762, 72
\bibitem[Vera-Ciro et al.(2011)]{ver-etal11} 
Vera-Ciro, C.~A., Sales, L.~V., Helmi, A., et al.\ 2011, \mnras, 416, 1377
\bibitem[Wang et al.(2014)]{wan-eatl14} 
Wang, X., Szalay, A., Arag{\'o}n-Calvo, M.~A., et al.\ 2014, \apj, 793, 58
\bibitem[Wang et al.(2018)]{wan-etal18} 
Wang, P., Guo, Q., Kang, X., et al.\ 2018, \apj, 866, 138
\bibitem[Welker et al.(2020)]{wel-etal20} 
Welker, C., Bland-Hawthorn, J., Van de Sande, J., et al.\ 2020, \mnras, 491, 2864
\bibitem[White(1984)]{whi84} 
White, S.~D.~M.\ 1984, \apj, 286, 38
\bibitem[Zhu \& Feng(2017)]{ZF17} 
Zhu, W. \& Feng, L.-L.\ 2017, \apj, 838, 21

\end{thebibliography}
\end{document}